\documentstyle[preprint,aps]{revtex}
\begin{document}
\draft

\title {Long-range behavior of the optical potential for
the elastic scattering of charged composite particles}
\author {E. O. Alt and A. M. Mukhamedzhanov $^{\dag}$}

\address
{Institut f\"ur Physik, Universit\"at Mainz, D-55099
Mainz, Germany}
\date{\today}
\maketitle
\begin{abstract}
The asymptotic behavior of the optical potential, describing
elastic scattering of a charged particle $\alpha$ off a bound
state of two charged, or one charged and one neutral,
particles at small momentum transfer $\Delta_{\alpha}$ or
equivalently at large intercluster distance $\rho_{\alpha}$,
is investigated within the framework of the exact three-body
theory. For the three-charged-particle Green function that
occurs in the exact expression for the optical potential, a
recently derived expression, which is appropriate
for the asymptotic region under consideration, is used.
We find that for arbitrary values of the energy parameter
the non-static part of the optical
potential behaves for $\Delta_{\alpha} \rightarrow 0$ as
$C_{1}\Delta_{\alpha} + o\,(\Delta_{\alpha})$. From this we
derive for the Fourier transform of its on-shell restriction
for $\rho_{\alpha} \rightarrow \infty$ the behavior
$-a/2\rho_{\alpha}^4 + o\,(1/\rho_{\alpha}^4)$, i.e., dipole
or quadrupole terms do
not occur in the coordinate-space asymptotics. This result
corroborates the standard one, which is obtained by
perturbative methods. The general, energy-dependent expression
for the dynamic polarisability $C_{1}$ is derived; on the
energy shell it reduces to the conventional polarisability
$a$ which is independent of the energy. We emphasize that
the present derivation is {\em non-perturbative}, i.e.,
it does not make use of adiabatic or
similar approximations, and is valid for energies {\em below
as well as above the three-body dissociation threshold}.
\end {abstract}
\pacs{03.65.Nk, 34.40.+n, 24.10.-i, 21.45.+v}

\newpage
\section {Introduction}

The incorporation of the long-ranged Coulomb interaction
in the theoretical description of reactions between charged
composite particles constitutes a problem of long-standing
interest, due to its great importance in atomic and
nuclear physics. As a prototype we consider the elastic
scattering of a charged particle off a cluster composed of
two charged (or one charged and one neutral) particles.

Formally, such a reaction can be described in terms of
a single-channel Lippmann-Schwinger-type integral equation
for the elastic scattering amplitude, in which the so-called
optical potential occurs (see, e.g., \cite{brans70,joach75}).
Although the latter is structurally too complicated for
practical purposes (besides being non-local, energy-dependent,
and complex above the threshold for the opening
of the lowest inelastic channel, it contains the three-body
Green function which is actually the object whose calculation
is attempted), its very existence is of considerable
significance for many purposes.

An alternative formulation is based on the
exact three-particle equations \cite{fad60}, suitably
generalized to accommodate the long-ranged Coulomb potentials
either in coordinate \cite{merk80}, or in
momentum space \cite{ves70,ves78,asz78,a78,as80}. The latter
approach leads to coupled multichannel Lippmann-Schwinger-type
equations whose solution yield simultaneously all
two-fragment, i.e. the (in-)elastic and rearrangement,
amplitudes (and with an additional quadrature
also the dissociation amplitudes). For three charged
particles, the effective potentials occurring therein are
again too complicated to be used presently for practical
calculations (except with drastic approximations). But, if
only two of the three particles are charged, some version
\cite{asz78} of these equations becomes
manageable, and has in fact already
been applied successfully to the calculation of
elastic scattering and breakup observables in the
proton-deuteron system \cite{asz85,ar94}.

Due to the differences in the physical pictures on which
these two formulations are based, the expressions for the
optical potential and for the effective potential as used
in conventional applications of the
three-body theory, differ appreciably (indeed, in ref.
\cite{gs69} it has been shown that the former can be
derived within the framework of the three-body theory
only under very special and unusual assumptions). It is
therefore of general interest to
gain insight into such properties of effective
potentials, which do not depend on the specific
details of the formulation, and are therefore common
to both of them. One important
example is the behavior for large separation of the
colliding particles which should be
similar in both cases, in contrast to the short-range
behavior which in general will be rather different.

In the present investigation we concentrate on the
optical potential. It is known to simplify considerably
when the distance between the colliding particles
goes to infinity, or equivalently when the momentum
transfer goes to zero. In fact, its large-distance behavior has
been extracted long time ago by various perturbative methods
below \cite{mw59,ss77}, and recently also above \cite{uc89}
(cf. also \cite{mcc80,mcc82}) the dissociation threshold.
The result is
well-known: besides the static potential which comprises
the multipole contributions arising from the charge
distribution of the composite particle, the first
non-vanishing term of longest range is the (local)
polarisation potential (proportional to the inverse fourth
power of the distance between the two colliding bodies).
But in none of these investigations the reliability of the
approximate, or the convergence
of the perturbative, treatment has been studied.

However, precise knowledge of the asymptotic behavior of
the optical potential is of great interest, for
many reasons: not only does it constrain the construction of
(the long-range part of) model optical potentials for
practical applications; but it also provides
answers to crucial questions like, e.g., what kind of
effective-range expansion be adequate or what rate of
convergence of partial-wave sums is to be expected.

For these reasons, there have recently appeared several
attempts to provide non-perturbative derivations based
on three-body theory, both in coordinate
\cite{km88} and in momentum \cite{ksz85,ks87,ksp87,ks89,zep90}
space. Indeed, it was found that the above mentioned,
approximately derived long-range behavior could be
recovered. However, not only are these investigations lacking
the rigour, and often also the generality, desirable in view
of the significance of the result. What is much more restrictive
is that these proofs could be established only for energies
{\em below the three-body dissociation threshold}.

In the present paper we give, based on the three-body integral
equations approach in momentum space, a {\em non-perturbative}
derivation of the behavior of the optical potential as the
momentum transfer goes to zero, which is valid {\em both below
and above} the three-body dissociation threshold, and which
does not suffer from the shortcomings of the previous
attempts. Of course, this momentum-transfer behavior
determines the large-distance behavior in coordinate space.
At the same time we deduce the general
(energy-dependent) form of the induced dipole
polarisability. For this purpose, use is made of the
recently developed asymptotic form of the wave function for
three charged particles in the continuum or equivalently
of the asymptotic expression for the three-charged-particle
Green function \cite{am92}, that applies to
just that region of the three-body configuration space
which is relevant in the present context.

The paper is organized as follows. In Sec. \ref{sec2} we
briefly recapitulate the formulation of the optical potential
for elastic scattering from the $m$-th bound state
(either ground or excited state) of the composite particle
within the framework of the three-body theory. When writing
down explicitly its leading term in the limit of
vanishing momentum transfer, the above mentioned
asymptotic expression for the three-charged-particle
Green function is introduced. The behavior
of the optical potential as the momentum transfer
goes to zero is derived in Sec. \ref{sec3}. There we also
present the general expression of the induced dipole
polarisability. These results are then specialized to
on-shell scattering, both in momentum and in coordinate
space. The Summary contains a discussion of our achievements.
Several auxiliary results are collected in the Appendices.

We mention that throughout we use natural units, i.e.,
$\hbar = c = 1$. Furthermore, unit vectors will be denoted by
a hat, i. e., $\hat a \equiv {\rm {\bf {a}}}/a$.

\newpage
\section{The optical potential} \label{sec2}

\subsection{Definition of the optical potential}

Consider three distinguishable particles with masses
$m_{\nu}$ and charges $e_{\nu},\,\nu = 1,2,3$. We use
Jacobi coordinates:
${\rm {\bf {p}}}_{\alpha} \,({\rm {\bf {r}}}_{\alpha})$ is the
relative momentum (coordinate) between particles $\beta$ and
$\gamma$, and $\mu_{\alpha} =m_{\beta}m_{\gamma}/(m_{\beta} +
m_{\gamma})$ their reduced mass; ${\rm {\bf {q}}}_{\alpha} \,
({\mbox {\boldmath $\rho$}}_{\alpha})$ denotes the relative
momentum (coordinate) between
particle $\alpha$ and the center of mass of the pair $(\beta
\gamma)$, the corresponding reduced mass being defined as
$M_{\alpha} = m_{\alpha}(m_{\beta}+m_{\gamma})/(m_{\alpha}
+m_{\beta}+m_{\gamma})$.

The Hamiltonian of the three-body system is
\begin{equation}
H = H_0 + V = H_0 + \sum_{\nu=1}^{3} V_{\nu}, \label{h}
\end{equation}
with $H_0$ being the free three-body Hamiltonian, and
\begin{equation}
V_{\alpha} = V_{\alpha}^S + V_{\alpha}^C \label{va}
\end{equation}
the full interaction between particles $\beta$ and $\gamma$,
consisting of a short-range ($V_{\alpha}^S$) and a Coulombic
part,
\begin{equation}
V_{\alpha}^C({\rm{\bf{r}}}_{\alpha}) =
\frac{e_{\beta}e_{\gamma}}{r_{\alpha}}.
\label{vcoul}
\end{equation}

The three-body transition operator $U_{\alpha \alpha}(z)$
which describes elastic and inelastic scattering in channel
$\alpha$, satisfies the equation
\begin{equation}
U_{\alpha \alpha}(z) = \bar V_{\alpha} + \bar V_{\alpha}
G_{\alpha}(z) U_{\alpha \alpha}(z). \label{uaa}
\end{equation}
Herein,
\begin{equation}
\bar V_{\alpha}=\sum_{\nu} \bar \delta_{\nu \alpha} V_{\nu} =
\bar V_{\alpha}^S + \bar V_{\alpha}^C =\sum_{\nu}
\bar \delta_{\nu \alpha} V_{\nu}^S +
\sum_{\nu} \bar \delta_{\nu \alpha} V_{\nu}^C \label{vbar}
\end{equation}
is the channel interaction and
\begin{equation}
G_{\alpha}(z) = (z - H_{\alpha})^{-1} =
(z - H_0 - V_{\alpha})^{-1} \label{ga}
\end{equation}
the channel Green function. The conventional notation
$\bar \delta_{\nu \alpha} = 1-\delta_{\nu \alpha}$ for the
anti-Kronecker symbol has been used.

Let $|\psi_{\alpha m}\rangle$ be the (normalized) bound
state wave function (belonging to the binding energy
$\hat E_{\alpha m}$, which we assume to be non-degenerate)
of the pair $(\beta \gamma)_{\,m}$, where the index $m$
denotes the complete set of quantum numbers. The notation
is to indicate that we allow the pair to be either in
the ground or in some excited state. The free motion of
particle $\alpha$ relative to the center of mass of
$(\beta \gamma)_{\,m}$ is described by
the plane wave $|{\rm {\bf {q}}}_{\alpha}\rangle$. Then
the quantity
\begin{eqnarray}
{\cal T}_{\alpha m, \alpha m}({\rm {\bf {q}}}
_{\alpha}', {\rm {\bf {q}}}_{\alpha}; E) =
\langle {\rm {\bf {q}}}_{\alpha}'| {\cal T}_{\alpha m,
\alpha m}(E+i0) |{\rm {\bf {q}}}_{\alpha}\rangle, \label{tel}
\end{eqnarray}
with
\begin{eqnarray}
{\cal T}_{\alpha m, \alpha m}(z) =
\langle \psi_{\alpha m} | U_{\alpha \alpha}(z)
| \psi_{\alpha m} \rangle, \label{tel1}
\end{eqnarray}
is on the energy shell, i.e., for
\begin{eqnarray}
E = E_{\alpha m} \equiv \bar q_{\alpha}^{2}/2 M_{\alpha} +
\hat E_{\alpha m}, \quad q_{\alpha}' =
q_{\alpha} = \bar q_{\alpha}, \label{oes}
\end{eqnarray}
the physical amplitude amplitude for elastic scattering
of particle $\alpha$ off the bound state
$(\beta \gamma)_{\,m}$.

Equation (\ref{uaa}) by itself does not yet lead to the
desired Lippmann-Schwinger (LS)-type equation for the
effective-two-body elastic scattering operator
${\cal T}_{\alpha m, \alpha m}(z) $
since the spectral decomposition of $G_{\alpha}(z)$ contains
contributions not only from the $m$-th but also from all the
other bound states, and in particular also from the continuum
states, in subsystem $\alpha$. This goal is achieved,
e.g., by means of the Feshbach projection operator technique
\cite{f58}. Introduce the projector onto the target state
and its orthogonal complement,
\begin{equation}
P_{\alpha m}=|\psi_{\alpha m}\rangle \langle \psi_{\alpha m}|,
\qquad Q_{\alpha m} = 1- P_{\alpha m}.\label{pq}
\end{equation}
Then
\begin{eqnarray}
G_{\alpha}(z)&=& P_{\alpha m} G_{\alpha}(z) + Q_{\alpha m}
G_{\alpha}(z) \nonumber \\
&=& G_{\alpha}^P(z) + G_{\alpha}^Q(z), \label{gpq}
\end{eqnarray}
with
\begin{eqnarray}
\langle {\rm {\bf {q}}}_{\alpha}'| G_{\alpha}^P(z) | {\rm {\bf
{q}}}_{\alpha} \rangle = |\psi_{\alpha m}\rangle
\frac{\delta({\rm {\bf {q}}}_{\alpha}' - {\rm {\bf
{q}}}_{\alpha})}{z - q_{\alpha}^{2}/2
M_{\alpha} - \hat E_{\alpha m} }\langle \psi_{\alpha m}|, \label{gp}
\end{eqnarray}
represents a splitting of the channel Green function into a
term (\ref{gp}) which is separable (with respect to the
variables internal to subsystem $\alpha$), and a remainder.
Introducing the decomposition
(\ref{gpq}) in (\ref{uaa}) yields, e.g. through application
of the AGS reduction procedure \cite{ags67},
\begin{eqnarray}
{\cal T}_{\alpha m, \alpha m}(z) = {\cal V}_{\alpha m,
\alpha m}^{opt}(z) + {\cal V}_{\alpha m, \alpha m}^{opt}(z)
\frac{1}{z - {\rm {\bf {Q}}}_{\alpha}^{\,2}/2 M_{\alpha}
+ \hat E_{\alpha m}}{\cal T}_{\alpha m, \alpha m}(z).
\label{ie}
\end{eqnarray}
For convenience we have introduced the relative momentum
operator ${\rm {\bf {Q}}}_{\alpha} $ whose eigenvalue is
${\rm {\bf {q}}}_{\alpha}$. Equation (\ref{ie}) is the desired
one-channel operator LS equation for the
elastic scattering amplitude. The plane wave matrix
elements of potential operator
\begin{eqnarray}
{\cal V}_{\alpha m, \alpha m}^{opt}({\rm {\bf {q}}}_{\alpha}',
{\rm {\bf {q}}}_{\alpha};z)
&=& \langle{\rm {\bf {q}}}_{\alpha}'|\langle\psi_{\alpha m} |
\bar V_{\alpha} + \bar V_{\alpha} Q_{\alpha m}
\frac{1}{z - H_{\alpha} - Q_{\alpha m} \bar V_{\alpha}
Q_{\alpha m}} Q_{\alpha m} \bar V_{\alpha} | \psi_{\alpha m}
\rangle |{\rm {\bf {q}}}_{\alpha} \rangle \nonumber \\
&=& \langle{\rm {\bf {q}}}_{\alpha}'|\langle \psi_{\alpha m} |
\bar V_{\alpha} + \bar V_{\alpha}G^Q(z)
\bar V_{\alpha} |\psi_{\alpha m} \rangle
|{\rm {\bf {q}}}_{\alpha} \rangle \label{vopt}
\end{eqnarray}
are seen to coincide with the standard definition of the
optical potential. Here, the identity
\begin{eqnarray}
Q_{\alpha m}[z - H_{\alpha} - Q_{\alpha m} \bar V_{\alpha}
Q_{\alpha m}]^{-1} Q_{\alpha m} = Q_{\alpha m} G(z)
Q_{\alpha m} =:G^{Q}(z), \label{ident}
\end{eqnarray}
$G(z) = (z - H)^{-1}$ being the resolvent of the full
Hamiltonian $H$, has been used. Of course,
this expression for the optical potential can also be derived
directly from the Schr\"odinger equation. Its relation to the
effective potentials introduced in the three-body
theory has been described in \cite{gs69}.

The question of the solvability of an equation like
(\ref{ie}) depends crucially on the singular
behavior of the effective potential
${\cal V}_{\alpha m, \alpha m}^{opt}({\rm{\bf{q}}}_{\alpha}',
{\rm {\bf {q}}}_{\alpha};E)$
in the limit that the momentum transfer
\begin{equation}
{\rm {\bf {\Delta}}}_{\alpha} = {\rm {\bf {q}}}_{\alpha}'-
{\rm {\bf {q}}}_{\alpha} \label{mt}
\end{equation}
goes to zero. In leading order the latter is defined
entirely by the Coulombic part $\bar V_{\alpha}^C$ of the
channel interaction, i.e., it does depend neither on the
short-ranged part $\bar V_{\alpha}^S$ nor on the internal
interaction $V_{\alpha}$. Of course, a given behavior of the
optical potential for $\Delta_{\alpha}  \rightarrow 0$
reflects itself in a corresponding asymptotic behavior
in coordinate space.

\subsection{Explicit expression}

In this section we establish the explicit expression of the
optical potential which will be used in the following
investigation of its analytical behavior in the limit
$\Delta_{\alpha} \rightarrow 0$.

Let us rewrite (\ref{vopt}) as
\begin{eqnarray}
{\cal V}_{\alpha m, \alpha m}^{opt}({\rm{\bf{q}}}_{\alpha}',
{\rm{\bf{q}}}_{\alpha}; E) = {\cal V}_{\alpha m, \alpha m}
^{opt\,(1)}({\rm{\bf{q}}}_{\alpha}', {\rm{\bf{q}}}_{\alpha}) +
{\cal V}_{\alpha m, \alpha m}^{opt\,(2)}({\rm
{\bf{q}}}_{\alpha}', {\rm{\bf{q}}}_{\alpha}; E),
\label{mvopt}
\end{eqnarray}
with
\begin{eqnarray}
{\cal V}_{\alpha m, \alpha m}^{opt\,(1)}
({\rm{\bf{q}}}_{\alpha}',{\rm{\bf{q}}}_{\alpha}) &=&
\langle {\rm{\bf{q}}}_{\alpha}'|\langle \psi_{\alpha m} |
\bar V_{\alpha} |\psi_{\alpha m}\rangle|
{\rm{\bf{q}}}_{\alpha} \rangle, \label{vopt1}\\
{\cal V}_{\alpha m, \alpha m}^{opt\,(2)}
({\rm{\bf{q}}}_{\alpha}', {\rm{\bf{q}}}_{\alpha};E)
&=& \langle {\rm{\bf{q}}}_{\alpha}'|\langle \psi_{\alpha m} |
\bar V_{\alpha} G^Q(E + i0) \bar V_{\alpha} | \psi_{\alpha m}
\rangle | {\rm{\bf{q}}}_{\alpha}\rangle. \label{vopts}
\end{eqnarray}
The first term (\ref{vopt1}) on the r.h.s. of (\ref{mvopt}),
which is called the static potential, has for $\Delta_{\alpha}
\rightarrow 0$ only the trivial Coulomb-type singular
behavior
\begin{eqnarray}
{\cal V}_{\alpha m, \alpha m}^{opt\,(1)}
({\rm{\bf{q}}}_{\alpha}',{\rm{\bf{q}}}_{\alpha})
&\stackrel{\Delta_{\alpha} \rightarrow 0}{\approx}&
\langle {\rm{\bf{q}}}_{\alpha}'|\langle \psi_{\alpha m} |
\bar V_{\alpha}^C |\psi_{\alpha m}\rangle|
{\rm{\bf{q}}}_{\alpha} \rangle \nonumber \\
&\stackrel{\Delta_{\alpha} \rightarrow 0}{=}&
\frac{4\pi e_{\alpha}(e_{\beta}+e_{\gamma})}
{\Delta_{\alpha}^{2}} + \cdots \label{vopt1s}
\end{eqnarray}
(the dots are to indicate terms $ \sim \Delta_{\alpha}^{-1}$
and $\sim \ln \Delta_{\alpha}$ which may arise, e.g., if the
target state is not spherically symmetric). It
can be taken care of in the LS equation (\ref{ie}) in
the usual manner (see, e.g., \cite{as80}).

Consequently the decisive question concerns the behavior of
${\cal V}_{\alpha m, \alpha m}^{opt\,(2)}
({\rm{\bf{q}}}_{\alpha}',{\rm{\bf{q}}}_{\alpha}; E)$ in
the limit of zero momentum transfer. In lowest order
perturbation theory, which in the
present language is equivalent to approximating in
(\ref{vopts}) the full three-body Green function $G(z)$ by its
lowest order approximation $G_{\alpha}(z)$,
and by taking into account that only the Coulombic part
$\bar V_{\alpha}^C$ of the channel potential
contributes to the leading singular behavior, one has
on the energy shell the well-known result
\begin{eqnarray}
{\cal V}_{\alpha m, \alpha m}^{opt\,(2)}({\rm{\bf{q}}}_{\alpha}',
{\rm{\bf{q}}}_{\alpha}; E_{\alpha m})
\stackrel{ \Delta_{\alpha}\rightarrow 0}{\approx}
\langle {\rm{\bf{q}}}_{\alpha}'|\langle \psi_{\alpha m} |
\bar V_{\alpha}^C G_{\alpha}^Q(E_{\alpha m} + i0)
\bar V_{\alpha}^C |
\psi_{\alpha m}\rangle | {\rm{\bf{q}}}_{\alpha}\rangle
\stackrel{\Delta_{\alpha} \rightarrow 0}{\sim}
\Delta_{\alpha}, \label{as0}
\end{eqnarray}
or in coordinate space
\begin{eqnarray}
{\cal V}_{\alpha m, \alpha m}^{opt\,(2)}
({\mbox{\boldmath $\rho$}}_{\alpha})
\stackrel{\rho_{\alpha} \rightarrow \infty}{\approx}
- \frac{a}{2 \rho_{\alpha}^4}, \label{as1}
\end{eqnarray}
for all energies. The factor $a$ represents the polarizability
of the
composite particle. However, it is of outstanding importance
to know whether this fundamental result holds also for the
exact optical potential (\ref{vopts}), i.e., even after all
terms of the perturbation expansion of $G(z)$ are summed up.
In fact, for energies below the three-body dissociation
threshold, i.e. for $E < 0$, it has been suggested in
\cite{km88,ksz85,ks87,ksp87,ks89,zep90} that (\ref{as0})
remains true also for the exact expression (\ref{vopts}).
It is the purpose of the present investigation to derive its
behavior for arbitrary $E$, that is in particular also for
$E > 0$.

Let us write down the matrix element (\ref{vopts})
explicitly in the coordinate-space representation,
\begin{eqnarray}
{\cal V}_{\alpha m, \alpha m}^{opt\,(2)}
({\rm{\bf{q}}}_{\alpha}', {\rm{\bf{q}}}_{\alpha}; E)
&=& \sum_{\nu}\bar \delta_{\nu \alpha}  \,
\sum_{\mu}\bar \delta_{\mu \alpha}
\int d {\mbox {\boldmath $\rho$}}_{\alpha}'\,d {\rm{\bf{r}}}_{\alpha}'\,
d {\mbox {\boldmath $\rho$}}_{\alpha}\,d {\rm{\bf{r}}}_{\alpha}\,
e^{-i {\rm{\bf{q}}}_{\alpha}' \cdot
{\mbox {\boldmath $\rho$}}_{\alpha}'} \nonumber \\
&&\times \psi_{\alpha m}^{*}({\rm{\bf{r}}}_{\alpha}')
V_{\mu}(\epsilon_{\alpha \mu} {\mbox {\boldmath $\rho$}}_{\alpha}' -
\lambda_{\mu}{\rm{\bf{r}}}_{\alpha}')
G^{Q}({\mbox {\boldmath $\rho$}}_{\alpha}',
{\rm{\bf{r}}}_{\alpha}';{\mbox {\boldmath $\rho$}}_{\alpha},
{\rm{\bf{r}}}_{\alpha}; E+i0)\nonumber \\
&&\times
V_{\nu}(\epsilon_{\alpha \nu}
{\mbox{\boldmath$\rho$}}_{\alpha} -
\lambda_{\nu}{\rm{\bf{r}}}_{\alpha})
\psi_{\alpha m}({\rm{\bf{r}}}_{\alpha})
 e^{-i {\rm{\bf{q}}}_{\alpha} \cdot
{\mbox {\boldmath $\rho$}}_{\alpha}}.
\label{vcoo}
\end{eqnarray}
For convenience, we have introduced the antisymmetric symbol
$\epsilon_{\beta \alpha} = -\epsilon_{\alpha \beta}$, with
$\epsilon_{\alpha \beta} = +1$ for $(\alpha,\beta)$ being a
cyclic ordering of $(1,2,3)$, and the mass ratios
$\lambda_{\nu} = m_{\nu}/(m_{\beta} +m_{\gamma})$,
for $\nu = \beta, \gamma$.

To further evaluate expression (\ref{vcoo}), the spectral
decomposition of the kernel of $G^{Q}(E+i0)$ must be inserted
and then the integrations over ${\rm{\bf{r}}}_{\alpha},
{\mbox {\boldmath $\rho$}}_{\alpha}, {\rm{\bf{r}}}_{\alpha}'$
and ${\mbox{\boldmath$\rho$}}_{\alpha}'$ must be carried out.
This requires in principle the knowledge of all the solutions
of the full three-body Schr\"odinger equation
\begin{equation}
\left\{ E - T_{{\rm{\bf{r}}}_{\alpha}} -
T_{{\mbox {\boldmath $\rho$}}_{\alpha}} - \sum_{\nu = 1}^{3}
V_{\nu} \right\} \Psi^{(+)} ({\rm{\bf{r}}}_{\alpha},
{\mbox {\boldmath $\rho$}}_{\alpha}) = 0 ,\label{se}
\end{equation}
whether they describe states with asymptotically two or three
unbound particles. Only the states corresponding to the
discrete spectrum of $H$, i.e. three-body bound states, are
not needed since they do not contribute to a singular
behavior of ${\cal V}_{\alpha m, \alpha m}^{opt\,(2)}$ in
this limit (the reason being simply that they yield only terms
which are separable in the incoming and outgoing momenta
and thus do not depend on the momentum transfer at all). Here,
$T_{{\rm{\bf{r}}}_{\alpha}}$ is the
kinetic energy operator for the relative motion of
particles $\beta$ and $\gamma$, and
$T_{{\mbox {\boldmath $\rho$}}_{\alpha}}$ the one for the
motion of particle $\alpha$ relative to the center of mass of
the pair $(\beta \gamma)_{\,m}$. They are defined as
 \begin{equation}
T_{{\rm{\bf{r}}}_{\alpha}} = -
\frac {\Delta_{{\rm{\bf{r}}}_{\alpha}}}{2\mu_{\alpha}},
\qquad T_{{\mbox {\boldmath $\rho$}}_{\alpha}} =
- \frac {\Delta_{{\mbox {\boldmath $\rho$}}_{\alpha}}}
{ 2M_{\alpha}}.\label{ekina}
\end{equation}

The solutions of (\ref{se}) are not known, but also not really
needed. For, we are interested only in the
behavior of ${\cal V}_{\alpha m, \alpha m}^{opt\,(2)}$ for
$\Delta_{\alpha} \rightarrow 0$.
Below it will be shown that in this limit the leading,
${\Delta}_{\alpha}$-dependent term is defined by
the singularities resulting
from the divergence of the integrals in (\ref{vcoo}) over
${\mbox {\boldmath $\rho$}}_{\alpha}$ for
${\rho}_{\alpha}\to \infty$, and over
${\mbox {\boldmath $\rho$}}_{\alpha}'$ for ${\rho}_{\alpha}'
\to \infty$. (Compare the discussion for an analogous
two-body problem in ref. \cite{am94}). The integrals over
${\rm{\bf{r}}}_{\alpha}$ and ${\rm{\bf{r}}}_{\alpha}'$ do
not induce a singular behavior. For, the
wave functions $\psi_{\alpha m}({\rm{\bf{r}}}_{\alpha})$
and $\psi_{\alpha m}^{*}({\rm{\bf{r}}}_{\alpha}')$ of the
incoming and outgoing bound pair $(\beta \gamma)_{\,m}$,
respectively, practically confine
the region of integration over the magnitudes of the
subsystem-internal variables
${\rm{\bf{r}}}_{\alpha}$ and ${\rm{\bf{r}}}_{\alpha}'$
to values ${r}_{\alpha}, {r}_{\alpha}' \leq
\kappa_{\alpha m}^{-1}$, where $\kappa_{\alpha m} = \sqrt{2
\mu_{\alpha} |\hat E_{\alpha m}|}$. For this reason,
when investigating the behavior of the leading term of
${\cal V}_{\alpha m, \alpha m}^{opt\,(2)}$ in the limit
$\Delta_{\alpha} \rightarrow 0$ it suffices to know the Green
function $G^{Q}({\mbox {\boldmath$\rho$}}_{\alpha}',
{\rm{\bf{r}}}_{\alpha}';{\mbox{\boldmath $\rho$}}_{\alpha},
{\rm{\bf{r}}}_{\alpha}; E+i0)$ in the asymptotic region
\begin{equation}
\Omega_{\alpha} \cap \Omega_{\alpha}', \label{asr}
\end{equation}
with
\begin{eqnarray}
\Omega_{\alpha}: \quad r_{\alpha}/\rho_{\alpha} \to 0, \quad
\rho_{\alpha} \to \infty, \label{om}\\
\Omega_{\alpha}': \quad r_{\alpha}'/\rho_{\alpha}' \to 0,
\quad \rho_{\alpha}' \to \infty. \label{oms}
\end{eqnarray}

In $\Omega_{\alpha}$ the Hamiltonian $H$ takes the form
\begin{equation}
H \stackrel{\Omega_{\alpha}}{\longrightarrow}
H^{as}_{\alpha} = T_{{\rm{\bf{r}}}_{\alpha}} +
T_{{\mbox {\boldmath $\rho$}}_{\alpha}} + V_{\alpha}({\rm{\bf{r}}}_{\alpha})
+ v^{C}_{\alpha}({\mbox {\boldmath $\rho$}}_{\alpha}),\label{hasa}
\end{equation}
where
\begin{equation}
v^{C}_{\alpha}({\mbox {\boldmath $\rho$}}_{\alpha}) =
\lim_{\rho_{\alpha} \rightarrow \infty,\,
r_{\alpha}/\rho_{\alpha} \rightarrow 0}
\left\{ V_{\beta}(\epsilon_{\alpha \beta} {\mbox {\boldmath $\rho$}}_{\alpha}
-\lambda_{\beta}{\rm{\bf{r}}}_{\alpha}) +
V_{\gamma}(\epsilon_{\alpha \gamma} {\mbox {\boldmath $\rho$}}_{\alpha}
-\lambda_{\gamma}{\rm{\bf{r}}}_{\alpha}) \right\} =
\frac{e_{\alpha}\,(e_{\beta} + e_{\gamma})}
{\rho_{\alpha}} \label{vacm}
\end{equation}
is the Coulomb potential between the charge $e_{\alpha}$ of
particle $\alpha$ and the total charge $(e_{\beta}+e_{\gamma})$
of the particles $\beta$ and $\gamma$ concentrated in their
center of mass. Because of this property $v^{C}_{\alpha}
({\mbox {\boldmath $\rho$}}_{\alpha})$ is
termed `center-of-mass Coulomb potential for channel
$\alpha$'. Note that in the region $\Omega_{\alpha}$ we could
neglect in (\ref{vacm}) the short-range interactions between
particles $\alpha$ and $\gamma$, and $\alpha$ and $\beta$,
completely, and terms $ \sim r_{\alpha}/\rho^{2}_{\alpha}$ in
the corresponding Coulombian parts.

The well-known class of solutions of the asymptotic
 Schr\"odinger equation
 \begin{equation}
\left\{E - H^{as}_{\alpha}\right\}\,\Psi^{(+)as}_
{{\rm {\bf {q}}}_{\alpha}^{0} m}
({\mbox {\boldmath $\rho$}}_{\alpha}
,{\rm{\bf{r}}}_{\alpha}) = 0, \label{seas1}
\end{equation}
belonging to the three-body energy
$E = {q_{\alpha}^{0}}^2/2 M_{\alpha} + \hat E_{\alpha m}$,
is given by
\begin{eqnarray}
\Psi^{(+)as}_{{\rm {\bf {q}}}_{\alpha}^{0} m}
({\mbox {\boldmath $\rho$}}_{\alpha},{\rm{\bf{r}}}_{\alpha}) =
\psi_{\alpha m}({\rm{\bf{r}}}_{\alpha})
\psi_{{\rm{\bf{q}}}_{\alpha}^{0}}^{(+)}
({\mbox {\boldmath $\rho$}}_{\alpha}).
\label{psid}
\end{eqnarray}
They consist of a product of the bound state wave function of the
pair $(\beta \gamma)_{\,m}$ introduced above, which satisfies the
two-body Schr\"odinger equation (we use the same symbols for
operators acting in the two- and in the three-particle space)
 \begin{equation}
\left\{ \hat E_{\alpha m}- T_{{\rm{\bf{r}}}_{\alpha}} -
V_{\alpha}({\rm{\bf{r}}}_{\alpha}) \right\}\,
\psi_{\alpha m}({\rm{\bf{r}}}_{\alpha}) = 0,\label{1a}
\end{equation}
and the `center-of-mass Coulomb scattering wave function', satisfying
 \begin{equation}
\left\{\frac{{q_{\alpha}^{0}}^2}{2 M_{\alpha}} -
T_{{\mbox {\boldmath $\rho$}}_{\alpha}} -
v^{C}_{\alpha}({\mbox {\boldmath $\rho$}}_{\alpha})\right\}\,
\psi^{(+)}_{{\rm{\bf{q}}}_{\alpha}^{0}}
({\mbox {\boldmath $\rho$}}_{\alpha}) = 0.\label{1b}
 \end{equation}
The solution of (\ref{1b}) is explicitly known,
\begin{eqnarray}
\psi^{(+)}_{{\rm{\bf{q}}}_{\alpha}^{0}}
({\mbox {\boldmath $\rho$}}_{\alpha}) =
e^{ i {\rm{\bf{q}}}_{\alpha}^{0}\cdot
{\mbox{\boldmath$\rho$}}_{\alpha}}
\bar N_{\alpha}^{0}F(-i \bar \eta_{\alpha}^{0},1;
i( q_{\alpha}^{0} \rho_{\alpha} - {\rm{\bf{q}}}_{\alpha}^{0}
\cdot {\mbox {\boldmath $\rho$}}_{\alpha})), \label{1c}
\end{eqnarray}
with
\begin{eqnarray}
\bar \eta_{\alpha}^{0} = \frac {e_{\alpha}
(e _{\beta}+ e_{\gamma})M_{\alpha}}{q_{\alpha}^{0}}, \quad
\bar N_{\alpha}^{0} = e^{- \pi \bar \eta_{\alpha}^{0}/2}\,
 \Gamma (1 + i \bar \eta_{\alpha}^{0}).\label{1dbar}
\end{eqnarray}
Here, $\Gamma(x)$ denotes the Gamma function, and $F(a,b;x)$
the confluent hypergeometric function \cite{abr70}. Note that
eventual bound state solutions of (\ref{1b}) for an attractive
center-of-mass Coulomb potential are of no interest in the
present context since they would correspond to a situation
with all three particles being confined (see below).

We also need the solution of the asymptotic Schr\"odinger
equation (\ref{seas1}) with all three particles asymptotically
in the continuum, i.e., the one which belongs to the
three-body energy $E = {q_{\alpha}^{0}}^2/2 M_{\alpha} +
{k_{\alpha}^{0}}^2/2 \mu_{\alpha}$. It will be called
asymptotic continuum solution of the Schr\"odinger equation in
$\Omega_{\alpha}$. This solution has been found in
\cite{am92}, and has the form
\begin{eqnarray}
\Psi^{(+)as}_{{\rm{\bf{q}}}_{\alpha}^{0}
{\rm{\bf{k}}}_{\alpha}^{0}}
({\mbox {\boldmath $\rho$}}_{\alpha},{\rm{\bf{r}}}_{\alpha}) =
\psi^{(+)}_{{\rm{\bf{k}}}_{\alpha}^{0}({\mbox {\boldmath $\rho$}}_{\alpha})}
({\rm {\bf{r}}}_{\alpha})\,e^{ i {\rm{\bf{q}}}_{\alpha}^{0} \cdot
{\mbox {\boldmath $\rho$}}_{\alpha}}
e^{i \eta_{\beta}^{0} ln(k_{\beta}^{0} \rho_{\alpha}-
\epsilon_{\alpha \beta} {\rm{\bf{k}}}_{\beta}^{0} \cdot
{\mbox {\boldmath $\rho$}}_{\alpha})}
e^{i \eta_{\gamma}^{0} ln(k_{\gamma}^{0} \rho_{\alpha}-
\epsilon_{\alpha \gamma} {\rm{\bf{k}}}_{\gamma}^{0} \cdot
{\mbox {\boldmath $\rho$}}_{\alpha})}.\label{psias}
\end{eqnarray}
The Coulomb parameters are given as
\begin{equation}
\eta_{\beta}^{0} = \frac {e_{\alpha} e_{\gamma}\mu_{\beta}}
{k_{\beta}^{0}}, \quad \eta_{\gamma}^{0} = \frac {e_{\alpha}
e_{\beta}\mu_{\gamma}}{k_{\gamma}^{0}},\label{cpar}
\end{equation}
and the momenta ${\rm{\bf{k}}}_{\beta}^{0}$ and
${\rm{\bf{k}}}_{\gamma}^{0}$ are defined as the usual linear
combinations of ${\rm{\bf{k}}}_{\alpha}^{0}$ and
${\rm{\bf{q}}}_{\alpha}^{0}$,
\begin{equation}
{\rm{\bf{k}}}_{\nu}^{0} = \epsilon_{\alpha \nu} \mu_{\alpha}
{\rm {\bf {q}}}_{\alpha}^{0}/M_{\nu} - \lambda_{\nu}
{\rm{\bf{k}}}_{\alpha}^{0}, \quad \mbox{with} \quad
\nu = \beta, \gamma. \label{lc}
\end{equation}
The wave function $\psi^{(+)}_{{\rm{\bf{k}}}_{\alpha}^{0}
({\mbox{\boldmath $\rho$}}_{\alpha})}
({\rm {\bf{r}}}_{\alpha})$ is the continuum solution of the
two-body-like Schr\"odinger equation
\begin{eqnarray}
\left\{ \frac {{k_{\alpha}^{0}}^{2}
({\mbox {\boldmath $\rho$}}_{\alpha})}
{2 \mu_{\alpha}} -T_{{\rm {\bf{r}}}_{\alpha}} -
V_{\alpha}({\rm {\bf{r}}}_{\alpha})\right\}
\psi^{(+)}_{{\rm{\bf{k}}}_{\alpha}^{0}
({\mbox {\boldmath $\rho$}}_{\alpha})}
({\rm {\bf{r}}}_{\alpha})= 0,\label{psis}
\end{eqnarray}
with
\begin{eqnarray}
{\rm{\bf{k}}}_{\alpha}^{0}({\mbox {\boldmath $\rho$}}_{\alpha})
&=& {\rm{\bf{k}}}_{\alpha}^0 +
\frac{{\rm{\bf{a}}}_{\alpha}({\hat {\mbox {\boldmath
$\rho$}}}_{\alpha})}{\rho_{\alpha}},\label{kloc} \\
{\rm{\bf{a}}}_{\alpha}({\hat {\mbox {\boldmath $\rho$}}}_{\alpha})
&=& - \sum_{\nu}\bar \delta_{\nu \alpha}
\eta_{\nu}^0 \lambda_{\nu} \frac{\epsilon_{\alpha \nu}
{\hat {\mbox {\boldmath $\rho$}}}_{\alpha} -
{\hat {\rm{\bf{k}}}}_{\nu}^0}{1 - \epsilon_{\alpha \nu}
{\hat {\mbox {\boldmath $\rho$}}}_{\alpha} \cdot
{\hat {\rm{\bf{k}}}}_{\nu}^0},\label{aa}
\end{eqnarray}
describing the relative motion of particles $\beta$ and
$\gamma$ with {\em local} energy $E_{\alpha}
({\mbox {\boldmath $\rho$}}_{\alpha}) = {k_{\alpha}^{0}}^{2}
({\mbox {\boldmath $\rho$}}_{\alpha})/2 \mu_{\alpha}$.
Thus it is, in fact, a three-body
wave function, the influence of the presence
of the third particle $\alpha$ however being confined to
a shift of the two-body relative momentum of particles
$\beta$ and $\gamma$ from its asymptotic
(for $\rho_{\alpha} \to \infty$, i.e. particle $\alpha$
is infinitely far apart) value
${\rm{\bf{k}}}_{\alpha}^{0}$ to the {\em local} value
${\rm{\bf{k}}}_{\alpha}^{0}({\mbox {\boldmath $\rho$}}_{\alpha})$. The latter
depends explicitly on the
distance of particle $\alpha$ from the center of mass of the
pair $(\beta \gamma)_{\,m}$ as a result of long-ranged three-body
correlations. Nevertheless, it is to be noted that, since
${\mbox {\boldmath $\rho$}}_{\alpha}$
enters the Schr\"odinger equation (\ref{psis}) only
parametrically via the local energy, its solutions
$\psi^{(+)}_{{\rm{\bf{k}}}_{\alpha}^{0}({\mbox
{\boldmath $\rho$}}_{\alpha})}({\rm {\bf{r}}}_{\alpha})$
and the genuine two-body bound state wave functions
$\psi_{\alpha m}({\rm{\bf{r}}}_{\alpha})$, which are
solutions of (\ref{1a}), are eigenfunctions of the same
Hamiltonian $\{T_{{\rm{\bf{r}}}_{\alpha}} +
V_{\alpha}({\rm{\bf{r}}}_{\alpha})\}$
to different eigenvalues, and are therefore orthogonal.

Let us add two comments.

1) It has to be pointed out that (\ref{psias}) is valid
in all of $\Omega_{\alpha}$, except for the so-called singular
directions which are defined by the conditions
$\epsilon_{\alpha \beta} {\hat {\rm{\bf{k}}}}_{\beta}^{0} \cdot
{\hat {\mbox {\boldmath $\rho$}}}_{\alpha} = 1$ and
$\epsilon_{\alpha \gamma} {\hat {\rm{\bf{k}}}}_{\gamma}^{0}\cdot
{\hat {\mbox {\boldmath $\rho$}}}_{\alpha} = 1$.
As was shown in \cite{am92}, in this whole region
$\Psi_{{\rm{\bf{q}}}_{\alpha}^{0}{\rm{\bf{k}}}_{\alpha}^{0}}
^{(+)as}({\mbox {\boldmath $\rho$}}_{\alpha},{\rm{\bf{r}}}_{\alpha})$
(i) satisfies the asymptotic Schr\"odinger equation
(\ref{seas1}), or equivalently the three-body
Schr\"odinger equation (\ref{se}) in $\Omega_{\alpha}$,
up to terms of the order $O(1/\rho_{\alpha}^2)$, viz.,
\begin{equation}
\left\{E - H^{as}_{\alpha} \right\} \,
\Psi^{(+)as}_{{\rm{\bf{q}}}_{\alpha}^{0}
{\rm{\bf{k}}}_{\alpha}^{0}}({\mbox {\boldmath $\rho$}}_{\alpha},
{\rm{\bf{r}}}_{\alpha}) = O\left(\frac{1}{\rho_{\alpha}^2}\right),
\label{seas}
\end{equation}
and (ii) represents the leading term in the asymptotic
expansion in $\Omega_{\alpha}$ of the full wave function
$\Psi^{(+)}_{{\rm{\bf{q}}}_{\alpha}^{0}{\rm{\bf{k}}}_{\alpha}^{0}}
({\mbox {\boldmath $\rho$}}_{\alpha},{\rm{\bf{r}}}_{\alpha})$,
which belongs to three asymptotically free
particles in the continuum with
Jacobi momenta ${\rm{\bf{k}}}_{\alpha}^{0}$ and
${\rm{\bf{q}}}_{\alpha}^{0}$, viz.,
\begin{equation}
\Psi^{(+)}_{{\rm{\bf{q}}}_{\alpha}^{0}{\rm{\bf{k}}}_{\alpha}^{0}}
({\mbox{\boldmath$\rho$}}_{\alpha}, {\rm{\bf{r}}}_{\alpha}) =
\Psi_{{\rm{\bf{q}}}_{\alpha}^{0}{\rm{\bf{k}}}_{\alpha}^{0}}
^{(+)as}({\mbox {\boldmath $\rho$}}_{\alpha},
{\rm{\bf{r}}}_{\alpha}) +O \left(\frac {1}{\rho_{\alpha}}\right).
\label{psias1}
\end{equation}
The next-to-leading term in the
expansion (\ref{psias1}) satisfies (\ref{seas}) in the next
orders. We remark that the second term in (\ref{psias1}),
which is of the order $O(1/\rho_{\alpha})$, includes
rescattering of the particles $\beta$ and $\gamma$ from
particle $\alpha$ \cite{merk80}. Its existence can easily be
deduced within the solvable model considered in \cite{am92}
(which corresponds to $V_{\alpha} = 0,\, m_{\alpha} = \infty$).

2) The asymptotic Schr\"odinger equation (\ref{seas1}) admits
also another (exact) solution, belonging to the same
energy $E = {q_{\alpha}^{0}}^2/2 M_{\alpha} +
{k_{\alpha}^{0}}^2/2 \mu_{\alpha}$ as (\ref{psias}), viz.
\begin{eqnarray}
\tilde \Psi^{(+)as}_{{\rm{\bf{q}}}_{\alpha}^{0}
{\rm{\bf{k}}}_{\alpha}^{0}}
({\mbox {\boldmath $\rho$}}_{\alpha},{\rm{\bf{r}}}_{\alpha}) =
\psi^{(+)}_{{\rm{\bf{k}}}_{\alpha}^{0}}({\rm{\bf{r}}}_{\alpha})
\psi^{(+)}_{{\rm{\bf{q}}}_{\alpha}^{0}}
({\mbox {\boldmath $\rho$}}_{\alpha}),\label{psias2}
\end{eqnarray}
where $\psi^{(+)}_{{\rm{\bf{k}}}_{\alpha}^{0}}
({\rm{\bf{r}}}_{\alpha})$ is again continuum solution of the
two-particle Schr\"odinger equation (\ref{psis}), but to the
energy ${k_{\alpha}^0}^2/2 \mu_{\alpha}$.
However, as shown in \cite{am92} this eigenfunction
of the asymptotic Hamiltonian $H_{\alpha}^{as}$ is not the
leading term in the asymptotic expansion in $\Omega_{\alpha}$
of the solution of the original Schr\"odinger equation
(\ref{se}), that is, it does not satisfy a relation like
(\ref{psias1}). Hence it is not to be used in the present
context (as was done mistakenly in \cite{km88}).

Consequently, the asymptotic form of the spectral
representation of the three-body Green function $G^Q(z)$,
which is valid in $\Omega_{\alpha} \cap \Omega_{\alpha}'$ and
is therefore appropriate for the investigation
of the singularity structure of ${\cal V}_{\alpha m, \alpha m}
^{opt\,(2)}$, has the form
\begin{eqnarray}
G^{Q}({\mbox {\boldmath $\rho$}}_{\alpha}',
{\rm{\bf{r}}}_{\alpha}';{\mbox {\boldmath $\rho$}}_{\alpha},
{\rm{\bf{r}}}_{\alpha}; E+i0)
&\stackrel{\Omega_{\alpha} \cap \Omega_{\alpha}'}
{\longrightarrow}&
G^{Q\,as}({\mbox {\boldmath $\rho$}}_{\alpha}',
{\rm{\bf{r}}}_{\alpha}';{\mbox {\boldmath $\rho$}}_{\alpha},
{\rm{\bf{r}}}_{\alpha}; E+i0) = \nonumber \\
&&\int \frac{d {\rm{\bf{q}}}_{\alpha}^{0}}{(2\pi)^{3}}\left\{
\sum_{n \neq m}\frac{\psi_{\alpha n}({\rm{\bf{r}}}_{\alpha}')
\psi_{{\rm{\bf{q}}}_{\alpha}^{0}}^{(+)}
({\mbox {\boldmath $\rho$}}_{\alpha}')
\psi_{{\rm{\bf{q}}}_{\alpha}^{0}}^{(+)*}
({\mbox {\boldmath $\rho$}}_{\alpha})\psi_{\alpha n}^{*}
({\rm{\bf{r}}}_{\alpha})}
{[E + i0 - {q_{\alpha}^{0}}^2/2 M_{\alpha} -
\hat E_{\alpha n}]}
\right. \nonumber \\ &&+ \left.
\int \frac{d {\rm{\bf{k}}}_{\alpha}^{0} }{(2\pi)^{3}}
\frac{\Psi_{{\rm{\bf{q}}}_{\alpha}^{0} {\rm{\bf{k}}}
_{\alpha}^{0}}^{(+)as}({\mbox {\boldmath $\rho$}}_{\alpha}',
{\rm{\bf{r}}}_{\alpha}'){\Psi_{{\rm{\bf{q}}}_{\alpha}^{0}
{\rm{\bf{k}}}_{\alpha}^{0}}^{(+)as*}}
({\mbox {\boldmath $\rho$}}_{\alpha},{\rm{\bf{r}}}_{\alpha})}
{[E + i0 -{q_{\alpha}^{0}}^2/2 M_{\alpha} -
{k_{\alpha}^{0}}^2/2 \mu_{\alpha}]} \right\}.
\label{gas}
\end{eqnarray}

One last point concerns the fact that in the second term
on the r.h.s. of (\ref{gas}) the
integrations over ${\rm{\bf{q}}}_{\alpha}^{0}$ and
${\rm{\bf{k}}}_{\alpha}^{0}$ extend also over the singular
directions, i.e., directions such that
$\epsilon_{\alpha \beta}{\hat {\rm{\bf{k}}}}_{\beta}^{0} \cdot
{\hat {\mbox {\boldmath $\rho$}}}_{\alpha} = 1$ and
$\epsilon_{\alpha \gamma}{\hat {\rm{\bf{k}}}}_{\gamma}^{0}\cdot
{\hat {\mbox {\boldmath $\rho$}}}_{\alpha} = 1$ holds.
But there, as mentioned above, the asymptotic
expansion (\ref{psias1}) is not valid. Thus, in principle
in the vicinity of these directions the exact expression of
the three-body wave function
$\Psi^{(+)}_{{\rm{\bf{q}}}_{\alpha}^{0}{\rm{\bf{k}}}_{\alpha}^{0}}
({\mbox {\boldmath $\rho$}}_{\alpha},{\rm{\bf{r}}}_{\alpha})$
must be used, which is
however unknown. Since the latter is normalized to
$\delta({\rm{\bf{q}}}_{\alpha}^{0} -
{\rm{\bf{q}}}_{\alpha}^{0'})\delta({\rm{\bf{k}}}_{\alpha}^{0}
-{\rm{\bf{k}}}_{\alpha}^{0'})$, it has to be integrable
everywhere, including the singular directions. The consequence
is that the contribution from
the infinitely small regions containing the singular
directions is infinitely small \cite{am94}. Thus, when looking
for the asymptotic behavior of ${\cal V}_{\alpha m,
\alpha m}^{opt\,(2)}({\rm {\bf {q}}}_{\alpha}',
{\rm {\bf {q}}}_{\alpha};E)$ for $\Delta_{\alpha}
\rightarrow 0$, in the main order we can approximate the exact
wave function $\Psi^{(+)}_{{\rm{\bf{q}}}_{\alpha}^{0}
{\rm{\bf{k}}}_{\alpha}^{0}}
({\mbox {\boldmath $\rho$}}_{\alpha},{\rm{\bf{r}}}_{\alpha})$
by $\Psi_{{\rm{\bf{q}}}_{\alpha}^{0}{\rm{\bf{k}}}_{\alpha}^{0}}^
{(+)as}({\mbox {\boldmath $\rho$}}_{\alpha},
{\rm{\bf{r}}}_{\alpha})$ in the whole integration region,
including the singular directions. Of course, some care must
be taken before using expression (\ref{psias}) in the singular
directions since, as follows from its definition (\ref{aa}),
${\rm{\bf{a}}}_{\alpha}({\hat{\mbox{\boldmath $\rho$}}}_{\alpha})$
diverges there. One possible remedy has been proposed in the
Appendix of ref. \cite{am92}. It consists in writing
${\rm {\bf {a}}}_{\alpha}({\hat {\mbox {\boldmath $\rho$}}}_{\alpha})/
{{\rho}_{\alpha}} \approx i \sum_{\nu= \beta, \gamma}
\epsilon_{\alpha \nu} \lambda_{\nu}
{\mbox {\boldmath $\nabla$}}_{{\mbox {\boldmath $\rho$}}_{\alpha}}
\ln F(-i\eta_{\nu},1; i(k_{\nu}\rho_{\alpha} -
\epsilon_{\alpha \nu} {\rm {\bf{k}}}_{\nu} \cdot
{\mbox {\boldmath $\rho$}}_{\alpha}))$.
The right-hand side has the virtue of remaining regular
when the singular directions are approached, while
away from them it coincides for $\rho_{\alpha} \to \infty$
in leading order with the right-hand side of (\ref{aa}).

Summarizing, higher order terms in the expansion
(\ref{psias1}) will not contribute to the leading term of
${\cal V}_{\alpha m, \alpha m}^{opt\,(2)}
({\rm {\bf {q}}}_{\alpha}',{\rm {\bf {q}}}_{\alpha} ;E)$
in the limit $\Delta_{\alpha} \rightarrow 0$, because its
behavior is defined by the divergence of the
integrals over ${\mbox {\boldmath $\rho$}}_{\alpha}'$ and
${\mbox {\boldmath $\rho$}}_{\alpha}$ in (\ref{vcoo}) for
${\rho}_{\alpha}',{\rho}_{\alpha} \to \infty$. And the
contribution from the infinitely small
neighbourhoods of the singular directions is infinitely small.
Thus we can use the representation (\ref{gas}) everywhere.

As discussed above, in $\Omega_{\alpha}$ the potentials
$V_{\beta}$ and $V_{\gamma}$ occurring in (\ref{vcoo})
can in the main order in $1/{\rho}_{\alpha}$ be approximated
by their Coulombic parts $V_{\beta}^C$ and $V_{\gamma}^C$.
The same holds true in $\Omega_{\alpha}'$. Thus, in the
limit ${\Delta}_{\alpha} \rightarrow 0$ one gets for the
leading, $\Delta_{\alpha}$-dependent part
${\cal V}_{\alpha m,\alpha m}^{opt\,(as)}$ of
the optical potential term
${\cal V}_{\alpha m,\alpha m}^{opt\,(2)}$:
\begin{eqnarray}
{\cal V}_{\alpha m, \alpha m}^{opt\,(2)}
({\rm{\bf{q}}}_{\alpha}', {\rm{\bf{q}}}_{\alpha}; E)
&\stackrel{{\Delta}_{\alpha} \rightarrow 0}{\approx}&
{\cal V}_{\alpha m, \alpha m}^{opt\,(as)}
({\rm{\bf{q}}}_{\alpha}', {\rm{\bf{q}}}_{\alpha}; E)
\nonumber \\ &=&
\sum_{\mu,\nu}\bar \delta_{\nu \alpha} \bar \delta_{\mu \alpha}
\int_{{\rho}_{\alpha}' \geq A}
d {\mbox {\boldmath $\rho$}}_{\alpha}'\,d {\rm{\bf{r}}}_{\alpha}'\,
\int_{{\rho}_{\alpha} \geq A}
d {\mbox {\boldmath $\rho$}}_{\alpha}\,d {\rm{\bf{r}}}_{\alpha}
\nonumber \\
&&\times \int \frac{d {\rm{\bf{q}}}_{\alpha}^{0}}{(2\pi)^{3}}
e^{-i {\rm{\bf{q}}}_{\alpha}' \cdot
{\mbox {\boldmath $\rho$}}_{\alpha}'}
\psi_{\alpha m}^{*}({\rm{\bf{r}}}_{\alpha}')
V_{\mu}^C(\epsilon_{\alpha \mu} {\mbox {\boldmath $\rho$}}_{\alpha}'
-\lambda_{\mu}{\rm{\bf{r}}}_{\alpha}') \nonumber \\
&&\times \left\{\sum_{n \neq m}\frac{\psi_{\alpha n}({\rm{\bf{r}}}_{\alpha}')
\psi_{{\rm{\bf{q}}}_{\alpha}^{0}}^{(+)}
({\mbox {\boldmath $\rho$}}_{\alpha}')
\psi_{{\rm{\bf{q}}}_{\alpha}^{0}}^{(+)*}
({\mbox {\boldmath $\rho$}}_{\alpha})
\psi_{\alpha n}^{*}({\rm{\bf{r}}}_{\alpha})}
{[E + i0 - {q_{\alpha}^{0}}^2/2 M_{\alpha} -
\hat E_{\alpha n}]}\right. \nonumber \\ && + \left.
\int \frac{d {\rm{\bf{k}}}_{\alpha}^{0} }{(2\pi)^{3}}
\frac{\Psi_{{\rm{\bf{q}}}_{\alpha}^{0} {\rm{\bf{k}}}
_{\alpha}^{0}}^{(+)as}({\mbox {\boldmath $\rho$}}_{\alpha}',
{\rm{\bf{r}}}_{\alpha}'){\Psi_{{\rm{\bf{q}}}_{\alpha}^{0}
{\rm{\bf{k}}}_{\alpha}^{0}}^{(+)as*}}
({\mbox {\boldmath $\rho$}}_{\alpha},{\rm{\bf{r}}}_{\alpha})}
{[E + i0 -{q_{\alpha}^{0}}^2/2 M_{\alpha} -
{k_{\alpha}^{0}}^2/2 \mu_{\alpha}]} \right\}\nonumber \\
&& \times V_{\nu}^C(\epsilon_{\alpha \nu}
{\mbox{\boldmath$\rho$}}_{\alpha}-
\lambda_{\nu}{\rm{\bf{r}}}_{\alpha})
\psi_{\alpha m}({\rm{\bf{r}}}_{\alpha})
e^{i {\rm{\bf{q}}}_{\alpha} \cdot {\mbox {\boldmath $\rho$}}
_{\alpha}}. \label{vcoo1}
\end{eqnarray}
The radius $A$, which defines the lower limits in the integrals
over the magnitudes of ${\mbox {\boldmath $\rho$}}_{\alpha}'$ and
${\mbox {\boldmath $\rho$}}_{\alpha}$, has to be chosen so large
that $A \gg 1/\kappa_{\alpha m}$, with $\kappa_{\alpha m}$ defined
previously. This condition ensures that we are allowed to use
the asymptotic expansion (\ref{psias1}) and to replace the full
two-body potentials by their Coulombic parts. These approximations
have already been introduced in (\ref{vcoo1}).

\newpage
\section{Behavior of the optical potential in the limit of
vanishing momentum transfer} \label{sec3}

In this section we will make use of techniques and results
developed for an analogous problem in
two-particle scattering in \cite{am94}. In
$\Omega_{\alpha}$ we can use the asymptotic expansion
\begin{equation}
V_{\nu}^C(\epsilon_{\alpha \nu}{\mbox{\boldmath $\rho$}}_{\alpha}
-\lambda_{\nu}{\rm{\bf{r}}}_{\alpha}) = \frac {e_{\tilde \nu}
e_{\alpha}}{\rho_{\alpha}} + \epsilon_{\alpha \nu}
\lambda_{\nu}\frac {e_{\tilde \nu}e_{\alpha}}{\rho_{\alpha}^2}
({\hat {\mbox {\boldmath $\rho$}}}_{\alpha}
\cdot {\rm{\bf{r}}}_{\alpha}) + O \left(\frac {r_{\alpha}^2}
{\rho_{\alpha}^3}\right), \quad \mbox{with} \quad
\nu \neq \tilde \nu = \beta, \gamma. \label{mp}
\end{equation}
For the following it proves useful to apply the familiar screening
to the Coulomb potentials which we choose to be of exponential
type in coordinate space, i.e., $V_{\alpha}^C({\rm{\bf{r}}}_{\alpha})
\rightarrow V_{\alpha}^C({\rm{\bf{r}}}_{\alpha}) e^{-\delta r_{\alpha}},
\,\delta > 0$. The zero-screening limit $\delta \to +0$ will
then be taken at the end.
Substitute (\ref{psias}) and (\ref{mp}) into (\ref{vcoo1}).
The orthogonality of the bound state wave function
$\psi_{\alpha m}({\rm{\bf{r}}}_{\alpha})$ to all the other
bound state wave functions $\psi_{\alpha n}
({\rm{\bf{r}}}_{\alpha})$ for $n \neq m$ and to the scattering
wave functions $\psi_{{\rm{\bf{k}}}_{\alpha}^{0}
({\mbox {\boldmath $\rho$}}_{\alpha})}^{(+)}
({\rm{\bf{r}}}_{\alpha})$ causes the vanishing of
the contribution of the first term of the
expansion (\ref{mp}). Thus one obtains in leading
order for ${\Delta}_{\alpha} \rightarrow 0$:
\begin{eqnarray}
{\cal V}_{\alpha m, \alpha m}^{opt\,(as)}
({\rm{\bf{q}}}_{\alpha}', {\rm{\bf{q}}}_{\alpha}; E)
&\stackrel{{\Delta}_{\alpha} \rightarrow 0}{\approx}&
\lim_{\delta \to +0}
\int \frac{d {\rm{\bf{q}}}_{\alpha}^{0}}{(2\pi)^{3}}
\int_{{\rho}_{\alpha}' \geq A}d {\mbox {\boldmath $\rho$}}_{\alpha}'
\int_{{\rho}_{\alpha} \geq A}d {\mbox {\boldmath $\rho$}}_{\alpha}
\frac{e^{-i ({\rm{\bf{q}}}_{\alpha}'-{\rm{\bf{q}}}_{\alpha}^0)
\cdot{\mbox {\boldmath $\rho$}}_{\alpha}'}}{{\rho}_{\alpha}^{'2}}
e^{-\delta {\rho}_{\alpha}'}\nonumber \\
&&\times \left\{ \tilde \psi_{{\rm{\bf{q}}}_{\alpha}^{0}}^{(+)}
({\mbox {\boldmath $\rho$}}_{\alpha}')\sum_{n \neq m}
\frac{D_{mn}(\hat {\mbox {\boldmath $\rho$}}_{\alpha}')
D_{nm}(\hat {\mbox {\boldmath $\rho$}}_{\alpha})}
{[E + i0 - {q_{\alpha}^{0}}^2/2 M_{\alpha} -\hat E_{\alpha n}]}
\tilde \psi_{{\rm{\bf{q}}}_{\alpha}^{0}}^{(+)*}
({\mbox {\boldmath $\rho$}}_{\alpha})
\right. \nonumber \\ && + \left.
\int \frac{d {\rm{\bf{k}}}_{\alpha}^{0} }{(2\pi)^{3}}
\prod_{\nu = \beta,\gamma} e^{i \eta_{\nu}^0 \ln (k_{\nu}^0
\rho_{\alpha}' - {\rm{\bf{k}}}_{\nu}^0 \cdot
{\mbox {\boldmath $\rho$}}_{\alpha}')}
\prod_{\sigma = \beta,\gamma} e^{-i \eta_{\sigma}^0
\ln (k_{\sigma}^0 \rho_{\alpha} - {\rm{\bf{k}}}_{\sigma}^0 \cdot
{\mbox {\boldmath $\rho$}}_{\alpha}) }
\right. \nonumber \\ &&\times \left.
\frac{D_{m {\rm{\bf{k}}}_{\alpha}^{0}
({\mbox {\boldmath $\rho$}}_{\alpha}')}
(\hat {\mbox {\boldmath $\rho$}}_{\alpha}')
D_{{\rm{\bf{k}}}_{\alpha}^{0}({\mbox {\boldmath $\rho$}}_{\alpha}) m}
(\hat {\mbox {\boldmath $\rho$}}_{\alpha})}
{[E + i0 -{q_{\alpha}^{0}}^2/2 M_{\alpha} -
{k_{\alpha}^{0}}^2/2 \mu_{\alpha}]} \right \}
e^{-\delta \rho_{\alpha}} \frac{e^{i ({\rm{\bf{q}}}_{\alpha}
-{\rm{\bf{q}}}_{\alpha}^0)\cdot
{\mbox {\boldmath $\rho$}}_{\alpha}}}{\rho_{\alpha}^2},
\label{vcoo3}
\end{eqnarray}
where we have introduced
\begin{eqnarray}
D_{nm}(\hat {\mbox {\boldmath $\rho$}}_{\alpha}) &=&
\epsilon_{\alpha \beta} e_{\alpha} (\lambda_{\beta}
e_{\gamma} - \lambda_{\gamma}e_{\beta})
\int d {\rm{\bf{r}}}_{\alpha}
\psi_{\alpha n}^{*}({\rm{\bf{r}}}_{\alpha})
({\hat {\mbox {\boldmath $\rho$}}}_{\alpha} \cdot
{\rm{\bf{r}}}_{\alpha}) \psi_{\alpha m}
({\rm{\bf{r}}}_{\alpha}), \label{dnm} \\
D_{{\rm{\bf{k}}}_{\alpha}^{0}
({\mbox {\boldmath $\rho$}}_{\alpha}) m}
(\hat {\mbox {\boldmath $\rho$}}_{\alpha}) &=&
\epsilon_{\alpha \beta} e_{\alpha} (\lambda_{\beta}
e_{\gamma} - \lambda_{\gamma}e_{\beta})
\int d {\rm{\bf{r}}}_{\alpha}
\psi^{(+)*}_{{\rm{\bf{k}}}_{\alpha}^{0}({\mbox
{\boldmath $\rho$}}_{\alpha})}({\rm {\bf{r}}}_{\alpha})
({\hat {\mbox {\boldmath $\rho$}}}_{\alpha} \cdot
{\rm{\bf{r}}}_{\alpha}) \psi_{\alpha m}
({\rm{\bf{r}}}_{\alpha}), \label{dkm} \\
D_{mn}(\hat {\mbox {\boldmath $\rho$}}_{\alpha})
&=& D_{nm}^{*}(\hat {\mbox {\boldmath $\rho$}}_{\alpha}),
\quad D_{m {\rm{\bf{k}}}_{\alpha}^{0}
({\mbox {\boldmath $\rho$}}_{\alpha})}
(\hat {\mbox {\boldmath $\rho$}}_{\alpha}) =
D_{{\rm{\bf{k}}}_{\alpha}^{0}
({\mbox {\boldmath $\rho$}}_{\alpha}) m}^{*}
(\hat {\mbox {\boldmath $\rho$}}_{\alpha}). \label{sym}
\end{eqnarray}
For convenience, the plane wave has been extracted in (\ref{vcoo3})
from the center-of-mass Coulomb wave function (\ref{1c}),
by writing $\psi^{(+)}_{{\rm{\bf{q}}}_{\alpha}^{0}}
({\mbox {\boldmath $\rho$}}_{\alpha}) =
e^{ i {\rm{\bf{q}}}_{\alpha}^{0}\cdot
{\mbox{\boldmath$\rho$}}_{\alpha}}
\tilde \psi^{(+)}_{{\rm{\bf{q}}}_{\alpha}^{0}}
({\mbox {\boldmath $\rho$}}_{\alpha})$.

Inspection of (\ref{vcoo3}) reveals that the
integrals over $\rho_{\alpha}'$ and $\rho_{\alpha}$ will
diverge for $\rho_{\alpha}', \rho_{\alpha} \rightarrow \infty$
if and only if
$({\rm {\bf {q}}}_{\alpha}'-{\rm {\bf {q}}}_{\alpha}^0)^2
\rightarrow - \delta^2$ and $({\rm {\bf {q}}}_{\alpha}-
{\rm {\bf {q}}}_{\alpha}^0)^2 \rightarrow - \delta^2$,
respectively. These divergences will give rise to singularites
in the integrand of the final
${\rm{\bf{q}}}_{\alpha}^{0}$-integration at the positions
\begin{equation}
({\rm {\bf {q}}}_{\alpha}'-{\rm {\bf {q}}}_{\alpha}^0)^2 +
\delta^2 = 0 \quad \mbox{and} \quad ({\rm {\bf {q}}}_{\alpha}-
{\rm {\bf {q}}}_{\alpha}^0)^2 + \delta^2 = 0. \label{s12}
\end{equation}
For $\delta \rightarrow +0$ we get from
(\ref{s12}) for the positions of the singularities
\begin{equation}
{\rm{\bf{q}}}_{\alpha}^{0} = {\rm {\bf {q}}}_{\alpha}', \quad
{\rm{\bf{q}}}_{\alpha}^{0} = {\rm {\bf {q}}}_{\alpha}. \label{ss}
\end{equation}
Proceeding then in close analogy to \cite{am94}, it will be
shown below that the coincidence of these singularities when
integrating over ${\rm{\bf{q}}}_{\alpha}^{0}$ gives rise to a
singularity of ${\cal V}_{\alpha m,\alpha m}^{opt\,(as)}$ in
the limit ${\Delta}_{\alpha} \rightarrow 0$ of the form
\begin{eqnarray}
{\cal V}_{\alpha m, \alpha m}^{opt\,(as)}
({\rm{\bf{q}}}_{\alpha}', {\rm{\bf{q}}}_{\alpha}; E)
&\stackrel{{\Delta}_{\alpha} \rightarrow 0}{\sim}&
\Delta_{\alpha} = \sqrt{{q_{\alpha}'}^2 + q_{\alpha}^2 - 2
q_{\alpha}'q_{\alpha} z} \nonumber \\
&\stackrel{q_{\alpha}' \to q_{\alpha}}{\sim}&
\sqrt{1-z}, \quad z = {\hat{\rm {\bf {q}}}}_{\alpha}'\cdot
{\hat{\rm {\bf {q}}}}_{\alpha}. \label{sing}
\end{eqnarray}
That is, ${\cal V}_{\alpha m, \alpha m}^{opt\,(as)}$
has a singularity in the forward direction at $z=1$ in
the $z$-plane which defines the
behavior of the leading term in the limit $z \to 1$, or
equivalently ${\Delta}_{\alpha} \rightarrow 0$.

{}From (\ref{ss}) we can conclude that in the limit
${\Delta}_{\alpha} \rightarrow 0$ the main contribution
to the integral over ${\rm{\bf{q}}}_{\alpha}^{0}$ comes
from the neighbourhood
\begin{equation}
{\rm{\bf{q}}}_{\alpha}^{0} \approx {\rm {\bf {q}}}_{\alpha}
\approx {\rm {\bf {q}}}_{\alpha}'. \label{ss1}
\end{equation}
In \cite{am94} this has been proved in the case of
two-body scattering, but it holds also for the three-body case
considered here. For, the additional integration over
${\rm{\bf{k}}}_{\alpha}^{0}$ appearing in the intermediate
three-body continuum state contribution in (\ref{vcoo3}) does
not influence this conclusion. Thus the leading singular
behavior of ${\cal V}_{\alpha m, \alpha m}^{opt\,(as)}$
is defined by the divergence of the integrals over
${\mbox {\boldmath $\rho$}}_{\alpha}$, over
${\mbox {\boldmath $\rho$}}_{\alpha}'$, and over
${\rm{\bf{q}}}_{\alpha}^{0}$.

\subsection{Contribution from two-fragment intermediate states}

Let us first consider that part of (\ref{vcoo3}) which is due
to the contribution from all subsystem bound states $n \neq m$
in the intermediate state,
\begin{eqnarray}
{\cal V}_{\alpha m, \alpha m\,(b)}^{opt\,(as)}
({\rm{\bf{q}}}_{\alpha}', {\rm{\bf{q}}}_{\alpha}; E)
&\stackrel{{\Delta}_{\alpha} \rightarrow 0}{\approx}&
\lim_{\delta \to +0}
\int \frac{d {\rm{\bf{q}}}_{\alpha}^{0}}{(2\pi)^{3}}
\int_{{\rho}_{\alpha}'\geq A}d{\mbox {\boldmath $\rho$}}_{\alpha}'
\int_{{\rho}_{\alpha} \geq A}d {\mbox {\boldmath $\rho$}}_{\alpha}
\frac{e^{-i {\rm{\bf{q}}}_{\alpha}' \cdot
{\mbox {\boldmath $\rho$}}_{\alpha}'}}{{\rho}_{\alpha}^{'2}}
e^{-\delta {\rho}_{\alpha}'}
\psi_{{\rm{\bf{q}}}_{\alpha}^{0}}^{(+)}
({\mbox {\boldmath $\rho$}}_{\alpha}') \nonumber \\
&& \times \sum_{n \neq m}
\frac{D_{mn}(\hat {\mbox {\boldmath $\rho$}}_{\alpha}')
D_{nm}(\hat {\mbox {\boldmath $\rho$}}_{\alpha})}
{[q_{\alpha n}^{2}/2 M_{\alpha} + i0 -
{q_{\alpha}^{0}}^2/2 M_{\alpha}]}
\psi_{{\rm{\bf{q}}}_{\alpha}^{0}}^{(+)*}
({\mbox {\boldmath $\rho$}}_{\alpha}) e^{-\delta \rho_{\alpha}}
\frac{e^{i {\rm{\bf{q}}}_{\alpha} \cdot
{\mbox{\boldmath$\rho$}}_{\alpha}}}{\rho_{\alpha}^2}.
\label{vcoo4}
\end{eqnarray}
Since the behavior of
${\cal V}_{\alpha m, \alpha m\,(b)}^{opt\,(as)}$ for
${\Delta}_{\alpha} \rightarrow 0$ is determined by
momenta characterized by (\ref{ss1}), the propagator
\begin{equation}
d_n(q_{\alpha}^{0}) = \left[E + i0 - {q_{\alpha}^{0}}^2
/2 M_{\alpha} -\hat E_{\alpha n}\right]^{-1} \label{dn}
\end{equation}
can be taken out from under the integral over
${\rm{\bf{q}}}_{\alpha}^{0}$ at the point
$q_{\alpha}^{0}= q_{\alpha}$, provided it is not singular
there. Let us write (\ref{dn}) as
\begin{eqnarray}
d_n(q_{\alpha}^{0}) = 2 M_{\alpha} /\left[q_{\alpha n}^{2} -
{q_{\alpha}^{0}}^2+ i0 \right], \label{dn2}
\end{eqnarray}
with (recall (\ref{oes}))
\begin{equation}
q_{\alpha n}^{2} = 2 M_{\alpha} (E- \hat E_{\alpha n}) =
{\bar q_{\alpha}}^2 + 2 M_{\alpha} (\hat E_{\alpha m} -
\hat E_{\alpha n}), \quad n \neq m. \label{qpole}
\end{equation}
If ${\rm{\bf{q}}}_{\alpha}^{0} = {\rm{\bf{q}}}_{\alpha}$, then
for on-shell scattering, i.e. for $q_{\alpha} =
\bar q_{\alpha}$, $d_n(q_{\alpha})$ is always finite. However,
for off-shell scattering ($q_{\alpha} \ne \bar q_{\alpha}$),
$d_n(q_{\alpha})$ has for $E - \hat E_{\alpha n} > 0$ a pole
at $q_{\alpha}=q_{\alpha n}$. This pole occurs both for
$E > 0$ and for $E < 0$, i.e. for
energies above and below the dissociation threshold.

Summarizing, for $q_{\alpha}^0=q_{\alpha} \neq q_{\alpha n}$,
$d_n(q_{\alpha}^0)$ is not singular and can therefore be taken
out from under the integral over ${\rm{\bf{q}}}_{\alpha}^{0}$
at the momentum $q_{\alpha}^{0}= q_{\alpha}$ when trying to
extract the behavior of ${\cal V}_{\alpha m, \alpha m}
^{opt\,(as)}$ in the limit ${\Delta}_{\alpha} \rightarrow 0$
in leading order \cite{am94}. Clearly this is no longer
allowed at the point $q_{\alpha} = q_{\alpha n}$ where the
propagator $d_n(q_{\alpha})$ has a pole.

Thus, if $q_{\alpha} \neq q_{\alpha n}$, we have in leading order
\begin{eqnarray}
{\cal V}_{\alpha m, \alpha m\,(b)}^{opt\,(as)}
({\rm{\bf{q}}}_{\alpha}', {\rm{\bf{q}}}_{\alpha}; E)
&\stackrel{{\Delta}_{\alpha} \rightarrow 0}{\approx}&
\sum_{n \neq m}
\frac{2 M_{\alpha}}{q_{\alpha n}^{2} - q_{\alpha}^2}
\lim_{\delta \to +0} \int_{{\rho}_{\alpha}' \geq A}
d {\mbox {\boldmath $\rho$}}_{\alpha}'
\int_{{\rho}_{\alpha} \geq A}d {\mbox {\boldmath $\rho$}}_{\alpha}
\frac{e^{-i {\rm{\bf{q}}}_{\alpha}' \cdot
{\mbox {\boldmath $\rho$}}_{\alpha}'}}{{\rho}_{\alpha}^{'2}}
e^{-\delta {\rho}_{\alpha}'}
D_{mn}(\hat {\mbox {\boldmath $\rho$}}_{\alpha}')\nonumber \\
&&\times D_{nm}(\hat {\mbox {\boldmath $\rho$}}_{\alpha})
 e^{-\delta \rho_{\alpha}}
\frac{e^{i {\rm{\bf{q}}}_{\alpha} \cdot
{\mbox {\boldmath $\rho$}}_{\alpha}}}{\rho_{\alpha}^2}
\int \frac{d {\rm{\bf{q}}}_{\alpha}^{0}}{(2\pi)^{3}}
\left\{ \psi_{{\rm{\bf{q}}}_{\alpha}^{0}}^{(+)}
({\mbox {\boldmath $\rho$}}_{\alpha}')
\psi_{{\rm{\bf{q}}}_{\alpha}^{0}}^{(+)*}
({\mbox {\boldmath $\rho$}}_{\alpha})
 \right \}.\label{vcoo5}
\end{eqnarray}
We can use the completeness relation
\begin{equation}
\int \frac{d {\rm{\bf{q}}}_{\alpha}^{0}}{(2\pi)^{3}}
\left\{ \psi_{{\rm{\bf{q}}}_{\alpha}^{0}}^{(+)}
({\mbox {\boldmath $\rho$}}_{\alpha}')
\psi_{{\rm{\bf{q}}}_{\alpha}^{0}}^{(+)*}
({\mbox {\boldmath $\rho$}}_{\alpha})
 \right \} = \delta({\mbox {\boldmath $\rho$}}_{\alpha}' -
{\mbox {\boldmath $\rho$}}_{\alpha}) - \sum_{\kappa}
\psi_{\kappa}({\mbox {\boldmath $\rho$}}_{\alpha}')
\psi_{\kappa}^{*}({\mbox {\boldmath $\rho$}}_{\alpha}),
\label{compl}
\end{equation}
for the solutions of (\ref{1b}). Of course, if the
center-of-mass Coulomb potential $v^{C}_{\alpha}
({\mbox {\boldmath $\rho$}}_{\alpha})$ is repulsive,
the sum over all bound states $\kappa$ with wave functions
$\psi_{\kappa}({\mbox {\boldmath $\rho$}}_{\alpha})$
is missing. Introducing (\ref{compl}) into (\ref{vcoo5})
we obtain
\begin{eqnarray}
{\cal V}_{\alpha m, \alpha m\,(b)}^{opt\,(as)}
({\rm{\bf{q}}}_{\alpha}', {\rm{\bf{q}}}_{\alpha}; E)
\stackrel{{\Delta}_{\alpha} \rightarrow 0}{\approx}
\sum_{n \neq m}
\frac{2 M_{\alpha}}{q_{\alpha n}^{2} - q_{\alpha}^2}
\int_{{\rho}_{\alpha} \geq A}d {\mbox {\boldmath $\rho$}}_{\alpha}
\frac{e^{-i {\rm {\bf {\Delta}}}_{\alpha} \cdot
{\mbox {\boldmath $\rho$}}_{\alpha}}}{\rho_{\alpha}^4}
|D_{nm}(\hat {\mbox {\boldmath $\rho$}}_{\alpha})|^2. \label{vas}
\end{eqnarray}
To arrive at this expression we have already taken into account
that, owing to the presence of the factor $\rho_{\alpha}^{-4}$
in the integrand, the integral in (\ref{vas}) converges, so
that the limit $\delta \rightarrow +0$ and the integration over
${\mbox {\boldmath $\rho$}}_{\alpha}$ could be interchanged.

In fact, the result (\ref{vas}) is valid both for repulsive
and attractive center-of-mass Coulomb potentials. For,
when deriving it for the latter case, one only has to
bear in mind that the bound state part in
(\ref{compl}), when inserted in (\ref{vcoo5}), gives rise
to terms in which the integrals over
${\mbox {\boldmath $\rho$}}_{\alpha}$ and
${\mbox {\boldmath $\rho$}}_{\alpha}'$ do not diverge
for $\rho_{\alpha}, \rho_{\alpha}' \to \infty$, due to
the exponential decay of the bound state wave functions,
and thus lead to nonsingular expressions.
Furthermore, each of these terms is
separable with respect to ${\rm{\bf{q}}}_{\alpha}$ and
${\rm{\bf{q}}}_{\alpha}'$; thus they can not contribute
at all to the momentum-transfer dependence of
${\cal V}_{\alpha m, \alpha m\,(b)}^{opt\,(as)}$
in the limit ${\Delta}_{\alpha} \rightarrow 0$.

In the case $q_{\alpha} = q_{\alpha n}$, the propagator
$d_n(q_{\alpha}^{0})$ can not be taken out from under the
integral over ${\rm{\bf{q}}}_{\alpha}^{0}$ at the point
$q_{\alpha}^{0}=q_{\alpha}$. This
implies that at the discrete points $q_{\alpha} =
q_{\alpha n}, \, n \neq m$, which are accessible in
off-shell scattering only, the limit
${\Delta}_{\alpha} \rightarrow 0$ in
${\cal V}_{\alpha m, \alpha m\,(b)}^{opt\,(as)}$
requires special considerations. This investigation is
deferred to Appendix \ref{appa} where it is shown that at
these discrete momenta
${\cal V}_{\alpha m, \alpha m\,(b)}^{opt\,(as)}$
remains regular in the limit $\Delta_{\alpha} \rightarrow 0$.

It is now an easy task to extract the behavior of (\ref{vas})
for ${\Delta}_{\alpha} \rightarrow 0$. To this end we write
(\ref{dnm}) in the form
\begin{eqnarray}
D_{nm}(\hat {\mbox {\boldmath $\rho$}}_{\alpha}) &=&
{\hat {\mbox {\boldmath $\rho$}}}_{\alpha} \cdot
{\rm{\bf{D}}}_{nm}
= \frac {i}{\rho_{\alpha}} \lim_{{\rm{\bf{p}}} \to 0}
({\rm{\bf{D}}}_{nm} \cdot {\mbox{\boldmath$\nabla$}}_{\rm{\bf{p}}})
e^{-i {\rm{\bf{p}}} \cdot {\mbox {\boldmath $\rho$}}_{\alpha}},
\label{dnm1}
\end{eqnarray}
with
\begin{eqnarray}
{\rm{\bf{D}}}_{nm}&=& \epsilon_{\alpha \beta} e_{\alpha}
(\lambda_{\beta}e_{\gamma} - \lambda_{\gamma}e_{\beta})
\int d {\rm{\bf{r}}}_{\alpha}
\psi_{\alpha n}^{*}({\rm{\bf{r}}}_{\alpha})
{\rm{\bf{r}}}_{\alpha} \psi_{\alpha m}
({\rm{\bf{r}}}_{\alpha}), \label{dnm3}
\end{eqnarray}
and use the symmetry property (\ref{sym}). Then
\begin{eqnarray}
{\cal V}_{\alpha m, \alpha m\,(b)}^{opt\,(as)}
({\rm{\bf{q}}}_{\alpha}', {\rm{\bf{q}}}_{\alpha}; E)
\stackrel{{\Delta}_{\alpha} \rightarrow 0}{\approx}
\sum_{n \neq m}
\frac{2 M_{\alpha}}{q_{\alpha n}^{2} - q_{\alpha}^2}
\lim_{{\rm{\bf{p}}},{\rm{\bf{p}}}' \to 0} ({\rm{\bf{D}}}_{mn}\cdot
{\mbox{\boldmath $\nabla$}}_{\rm{\bf{p}}'})
({\rm{\bf{D}}}_{nm}\cdot
{\mbox {\boldmath $\nabla$}}_{\rm{\bf{p}}}) J(\rm{\bf{u}}),
\label{vas1}
\end{eqnarray}
the integral $J(\rm{\bf{u}})$ being defined as
\begin{eqnarray}
J(\rm{\bf{u}}) = \int_{{\rho}_{\alpha} \geq A}
d {\mbox {\boldmath $\rho$}}_{\alpha}
\frac{e^{i {\mbox {\boldmath $\rho$}}_{\alpha} \cdot
{\rm{\bf{u}}}}}{\rho_{\alpha}^6} = \frac {4 \pi}{u}
\int_{A}^{\infty} d \rho_{\alpha} \frac {\sin u \rho_{\alpha}}
{\rho_{\alpha}^5}, \label{ju}
\end{eqnarray}
with ${\rm{\bf{u}}} = {\rm{\bf{p}}}' -
{\rm {\bf {\Delta}}}_{\alpha} - {\rm{\bf{p}}}$. This integral
is evaluated in Appendix \ref{appb}. Substituting its
asymptotic form (\ref{b3}) for $u \to 0$ (i.e.
${\rm{\bf{p}}}', {\rm{\bf{p}}}$, and
$\Delta_{\alpha}$ going to zero) into (\ref{vas1}) leads to
\begin{eqnarray}
{\cal V}_{\alpha m, \alpha m\,(b)}^{opt\,(as)}
({\rm{\bf{q}}}_{\alpha}',{\rm{\bf{q}}}_{\alpha};E)
\stackrel{{\Delta}_{\alpha} \rightarrow 0}{=}
C_{m}^{(b)} \Delta_{\alpha} + o\,(\Delta_{\alpha}), \label{vas3}
\end{eqnarray}
with
\begin{eqnarray}
C_{m}^{(b)} = -\frac{\pi^2}{2} \sum_{n \neq m}
\frac{M_{\alpha}}{q_{\alpha n}^{2} - q_{\alpha}^2}
\left[|{\rm{\bf{D}}}_{nm}|^2 + |\hat {\rm {\bf {\Delta}}}_{\alpha}
\cdot {\rm{\bf{D}}}_{nm}|^2 \right]. \label{cn}
\end{eqnarray}
We mention that the action of
${\mbox {\boldmath $\nabla$}}_{\rm{\bf{p}}'}$ and
${\mbox {\boldmath $\nabla$}}_{\rm{\bf{p}}}$ onto the term
$\sim u^2$ in the asymptotic form (\ref{b3}) for
$J(\rm{\bf{u}})$ yields in the
limit ${\rm{\bf{p}}},{\rm{\bf{p}}}' \to 0$ also a
$\Delta_{\alpha}$-independent term. The latter was, however,
omitted in (\ref{vas3}) since
${\cal V}_{\alpha m, \alpha m\,(b)}^{opt\,(as)}$ was defined
as the leading, $\Delta_{\alpha}$-dependent contribution from
the intermediate two-fragment states to the nontrivial part
${\cal V}_{\alpha m, \alpha m}^{opt\,(2)}$ of the optical
potential in the limit ${\Delta}_{\alpha} \rightarrow 0$.

For on-shell scattering ($q_{\alpha}'=q_{\alpha}=\bar
q_{\alpha}$) we have
\begin{equation}
(q_{\alpha n}^{2} - q_{\alpha}^2)/2 M_{\alpha} =
\hat E_{\alpha m} - \hat E_{\alpha n}, \quad
n \neq m, \label{ons}
\end{equation}
and hence (\ref{cn}) becomes
\begin{eqnarray}
C_{m}^{(b)} &=& -\frac{\pi^2}{4} \sum_{n \neq m}
\frac{1}{\hat E_{\alpha m} - \hat E_{\alpha n}}
\left[|{\rm{\bf{D}}}_{nm}|^2 + |\hat {\rm {\bf {\Delta}}}_{\alpha}
\cdot {\rm{\bf{D}}}_{nm}|^2 \right]. \label{cn1}
\end{eqnarray}
Thus, ${\cal V}_{\alpha m, \alpha m\,(b)}^{opt\,(as)}$
as given by (\ref{vas3}) and (\ref{cn1}), depends only
on the momentum transfer
${\rm {\bf {\Delta}}}_{\alpha}$, and not on the incoming
or outgoing momentum ${\rm{\bf{q}}}_{\alpha}$ and
${\rm{\bf{q}}}_{\alpha}'$ separately, or on the energy.
Consequently, on the energy shell we can write (\ref{vas}) as
\begin{eqnarray}
{\cal V}_{\alpha m, \alpha m\,(b)}^{opt\,(as)}
({\rm{\bf{q}}}_{\alpha}', {\rm{\bf{q}}}_{\alpha}; E_{\alpha m}+i0)
&\equiv& {\cal V}_{\alpha m, \alpha m\,(b)}^{opt\,(as)}
({\rm {\bf {\Delta}}}_{\alpha}) \nonumber \\
&\stackrel{{\Delta}_{\alpha} \rightarrow 0}{\approx}&
\int_{{\rho}_{\alpha} \geq A}d {\mbox {\boldmath $\rho$}}_{\alpha}
e^{-i {\rm {\bf {\Delta}}}_{\alpha} \cdot
{\mbox {\boldmath $\rho$}}_{\alpha}}
{\cal V}_{\alpha m, \alpha m\,(b)}^{opt\,(as)}
({\mbox {\boldmath $\rho$}}_{\alpha}), \label{vdelt}
\end{eqnarray}
where
\begin{eqnarray}
{\cal V}_{\alpha m, \alpha m\,(b)}^{opt\,(as)}
({\mbox {\boldmath $\rho$}}_{\alpha}) =
\left(\sum_{n \neq m}\frac{|D_{nm}
(\hat {\mbox {\boldmath $\rho$}}_{\alpha})|^2}{\hat E_{\alpha m} -
\hat E_{\alpha n}}\right) \frac{1}{\rho_{\alpha}^4}
\label{vascoo}
\end{eqnarray}
describes the asymptotics of ${\cal V}_{\alpha m,
\alpha m\,(b)}^{opt\,(as)}$ in the coordinate space. It is
a local potential, representing the contribution to the
polarisation potential
from all the intermediate two-fragment states.

\subsection{Contribution from intermediate three-particle
continuum states}

We now proceed to investigate the contribution
${\cal V}_{\alpha m, \alpha m\,(c)}^{opt\,(as)}$ of the
intermediate three-particle continuum states to the optical
potential, which is given by the integral part in the wavy
brackets of (\ref{vcoo3}), in the limit
${\Delta}_{\alpha} \rightarrow 0$:
\begin{eqnarray}
{\cal V}_{\alpha m, \alpha m\,(c)}^{opt\,(as)}
({\rm{\bf{q}}}_{\alpha}', {\rm{\bf{q}}}_{\alpha}; E)
&\stackrel{{\Delta}_{\alpha} \rightarrow 0}{\approx}&
\lim_{\delta \to +0}
\int \frac{d {\rm{\bf{q}}}_{\alpha}^{0}}{(2\pi)^{3}}
\int_{{\rho}_{\alpha}' \geq A}d {\mbox {\boldmath $\rho$}}_{\alpha}'
\int_{{\rho}_{\alpha} \geq A}d {\mbox {\boldmath $\rho$}}_{\alpha}
\frac{e^{-i ({\rm{\bf{q}}}_{\alpha}'- {\rm{\bf{q}}}_{\alpha}^{0})
\cdot{\mbox {\boldmath $\rho$}}_{\alpha}'}}{{\rho}_{\alpha}^{'2}}
e^{-\delta {\rho}_{\alpha}'} \nonumber \\
&&\times e^{-\delta \rho_{\alpha}}
\frac{e^{i ({\rm{\bf{q}}}_{\alpha}-{\rm{\bf{q}}}_{\alpha}^{0})
\cdot{\mbox {\boldmath $\rho$}}_{\alpha}}}{\rho_{\alpha}^2}
L({\rm {\bf{q}}}_{\alpha}^{0}; {\mbox {\boldmath $\rho$}}_{\alpha},
{\mbox {\boldmath $\rho$}}_{\alpha}'), \label{vcoo6}
\end{eqnarray}
with
\begin{eqnarray}
L({\rm {\bf{q}}}_{\alpha}^{0}; {\mbox {\boldmath $\rho$}}_{\alpha},
{\mbox {\boldmath $\rho$}}_{\alpha}')&=&
\int \frac{d {\rm{\bf{k}}}_{\alpha}^{0} }{(2\pi)^{3}}
\prod_{\nu = \beta,\gamma} e^{i \eta_{\nu}^0 \ln (k_{\nu}^0
\rho_{\alpha}' - {\rm{\bf{k}}}_{\nu}^0 \cdot
{\mbox {\boldmath $\rho$}}_{\alpha}')}
\prod_{\sigma = \beta,\gamma} e^{-i \eta_{\sigma}^0
\ln (k_{\sigma}^0\rho_{\alpha} - {\rm{\bf{k}}}_{\sigma}^0 \cdot
{\mbox {\boldmath $\rho$}}_{\alpha}) } \nonumber \\
&&\times \frac{D_{m {\rm{\bf{k}}}_{\alpha}^{0}
({\mbox {\boldmath $\rho$}}_{\alpha}')}
(\hat {\mbox {\boldmath $\rho$}}_{\alpha}')
D_{{\rm{\bf{k}}}_{\alpha}^{0}({\mbox{\boldmath$\rho$}}_{\alpha}) m}
(\hat {\mbox {\boldmath $\rho$}}_{\alpha})}
{[E + i0 -{q_{\alpha}^{0}}^2/2 M_{\alpha} -
{k_{\alpha}^{0}}^2/2 \mu_{\alpha}]}. \label{l}
\end{eqnarray}

Consider the integration over ${\rm{\bf{q}}}_{\alpha}^{0}$
in (\ref{vcoo6}). Since, as discussed above, the behavior of
${\cal V}_{\alpha m, \alpha m}^{opt\,(as)}$ for
${\Delta}_{\alpha} \rightarrow 0$ is defined by
momenta characterized by the restriction (\ref{ss1}),
the function $L({\rm {\bf {q}}}_{\alpha}^{0};
{\mbox {\boldmath $\rho$}}_{\alpha},
{\mbox {\boldmath $\rho$}}_{\alpha}')$ can be taken out
from under the integral over ${\rm {\bf{q}}}_{\alpha}^{0}$
at the momentum ${\rm {\bf{q}}}_{\alpha}^{0}=
{\rm {\bf{q}}}_{\alpha}$ provided it is nonsingular there. A
singularity of $L$ in the ${\rm {\bf {q}}}_{\alpha}^{0}$-plane,
which may be of concern to us in the present context, can be
generated in (\ref{l}) only by the pole of the propagator
\begin{equation}
d_{k_{\alpha}^{0}}(q_{\alpha}^{0})=
\left[E + i0 - {q_{\alpha}^{0}}^2/2 M_{\alpha} -
{k_{\alpha}^{0}}^2/2 \mu_{\alpha} \right]^{-1}
\label{dk}
\end{equation}
$k_{\alpha}^{0}-$plane. In fact, this pole induces in the
function $L({\rm {\bf {q}}}_{\alpha}^{0};
{\mbox {\boldmath $\rho$}}_{\alpha},
{\mbox {\boldmath $\rho$}}_{\alpha}')$ a singularity in
the $q_{\alpha}^{0}$-plane at $q_{\alpha}^{0}=
\sqrt{2 M_{\alpha}E}$ (the positive square root is singled
out by the infinitely small imaginary part $+i0$ in the
propagator). It is an end point singularity which results from the
coincidence of the propagator pole with the lower limit zero
of the integration over $k_{\alpha}^{0}$.

Hence one has to distinguish two cases. \\
(i) $q_{\alpha} \neq {\sqrt {2 M_{\alpha}E}}$, i.e.,
the function $L({\rm {\bf{q}}}_{\alpha}^{0};
{\mbox {\boldmath $\rho$}}_{\alpha},
{\mbox {\boldmath $\rho$}}_{\alpha}')$ is regular at the
momentum ${\rm {\bf {q}}}_ {\alpha}^{0}={\rm {\bf {q}}}_{\alpha}$
(this is obvious since the propagator pole at $k_{\alpha}^{0}=
\sqrt{2 \mu_{\alpha}(E - q_{\alpha}^{2}/2 M_{\alpha})} \neq 0$
does not coincide
with the lower limit $k_{\alpha}^{0}=0$ of the integration in
(\ref{l})). As a consequence, $L({\rm {\bf {q}}}_{\alpha}^{0};
{\mbox {\boldmath $\rho$}}_{\alpha},
{\mbox {\boldmath $\rho$}}_{\alpha}')$ can be taken out
from under the integral over ${\rm {\bf {q}}}_{\alpha}^{0}$ at
${\rm {\bf {q}}}_{\alpha}^{0}= {\rm {\bf {q}}}_{\alpha}$ when looking
for the behavior of ${\cal V}_{\alpha m, \alpha m\,(c)}^{opt\,(as)}$
in the limit ${\Delta}_{\alpha} \rightarrow 0$.
This holds true {\em a fortiori} for the propagator
$d_{k_{\alpha}^{0}}(q_{\alpha}^{0})$ in (\ref{l}). For
$q_{\alpha}^{0}=q_{\alpha}$, the latter can be written as
(recall (\ref{oes}))
\begin{eqnarray}
d_{k_{\alpha}^{0}}(q_{\alpha})=\left[\bar q_{\alpha}^2/2
M_{\alpha} -q_{\alpha}^2/2 M_{\alpha}-
{k_{\alpha}^{0}}^2/2 \mu_{\alpha} + \hat E_{\alpha m}
+i0 \right]^{-1}. \label{dk2}
\end{eqnarray}
Note that on the energy shell, i.e., for $q_{\alpha}=
\bar q_{\alpha}$, expression (\ref{dk2}) simplifies to
\begin{equation}
d_{k_{\alpha}^{0}}(q_{\alpha})=-
\left[|\hat E_{\alpha m}| + {k_{\alpha}^{0}}^2/2 \mu_{\alpha}
\right]^{-1} < 0, \label{dk1}
\end{equation}
that is, the propagator $d_{k_{\alpha}^{0}}(q_{\alpha}^{0})$
is nonsingular at $q_{\alpha}^0=q_{\alpha}=\bar q_{\alpha}$
for all $k_{\alpha}^0$. \\
(ii) $q_{\alpha}= \sqrt {2 M_{\alpha}E}$, which can happen
only off the energy shell for $E>0$. In this case
$L({\rm {\bf {q}}}_{\alpha}^{0}; {\mbox {\boldmath $\rho$}}_{\alpha},
{\mbox {\boldmath $\rho$}}_{\alpha}')$ cannot be taken out
from under the integral over ${\rm {\bf {q}}}_{\alpha}^{0}$
in (\ref{vcoo6}) at the point ${\rm {\bf {q}}}_{\alpha}^{0}=
{\rm {\bf {q}}}_{\alpha}$ because it is singular there.
A closer examination, which is deferred to Appendix \ref{appa},
reveals that in the limit $\Delta_{\alpha} \to 0$,
${\cal V}_{\alpha m, \alpha m\,(c)}^{opt\,(as)}$ is
$O(\Delta_{\alpha})$ if the Coulomb potential $V_{\alpha}^C$
is repulsive, and that it remains regular for an attractive
Coulomb potential $V_{\alpha}^C$. In other words, the
leading singular behavior is unaltered as compared to case (i).

Summarizing, in order to isolate the leading singular part of
${\cal V}_{\alpha m, \alpha m\,(c)}^{opt\,(as)}$ for
$q_{\alpha} \neq {\sqrt {2 M_{\alpha}E}}$,
we can take in (\ref{vcoo6}) the propagator
$d_{k_{\alpha}^{0}}(q_{\alpha}^{0})$, as well as all
other $q_{\alpha}^{0}$-depending factors none of which
becomes singular there, out from under the integral over
${\rm{\bf{q}}}_{\alpha}^{0}$ at the point $q_{\alpha}^0 =
q_{\alpha}$. This gives
\begin{eqnarray}
{\cal V}_{\alpha m, \alpha m\,(c)}^{opt\,(as)}
({\rm{\bf{q}}}_{\alpha}', {\rm{\bf{q}}}_{\alpha}; E)
&\stackrel{{\Delta}_{\alpha} \rightarrow 0}{\approx}&
\lim_{\delta \to +0}
\int \frac{d {\rm{\bf{k}}}_{\alpha}^{0}}{(2\pi)^{3}}
\int_{{\rho}_{\alpha}' \geq A}d {\mbox {\boldmath $\rho$}}_{\alpha}'
\int_{{\rho}_{\alpha} \geq A}d {\mbox {\boldmath $\rho$}}_{\alpha}
\frac{e^{-i {\rm{\bf{q}}}_{\alpha}'
\cdot{\mbox {\boldmath $\rho$}}_{\alpha}'}}{{\rho}_{\alpha}^{'2}}
e^{-\delta {\rho}_{\alpha}'} \nonumber \\
&&\times\prod_{\nu = \beta,\gamma} \left(\frac{ \bar k_{\nu}^0
\rho_{\alpha}' - \bar {\rm{\bf{k}}}_{\nu}^0 \cdot
{\mbox {\boldmath $\rho$}}_{\alpha}'}{\bar k_{\nu}^0
\rho_{\alpha} - \bar {\rm{\bf{k}}}_{\nu}^0 \cdot
{\mbox {\boldmath $\rho$}}_{\alpha}}\right)^{i \bar \eta_{\nu}^0 }
e^{-\delta \rho_{\alpha}} \frac{e^{i {\rm{\bf{q}}}_{\alpha}
\cdot{\mbox {\boldmath $\rho$}}_{\alpha}}}{\rho_{\alpha}^2} \nonumber \\
&&\times \frac{D_{m {\rm{\bf{k}}}_{\alpha}^{0}
({\mbox {\boldmath $\rho$}}_{\alpha}')}
(\hat {\mbox {\boldmath $\rho$}}_{\alpha}')
D_{{\rm{\bf{k}}}_{\alpha}^{0}({\mbox{\boldmath$\rho$}}_{\alpha}) m}
(\hat {\mbox {\boldmath $\rho$}}_{\alpha})}
{[E + i0 - q_{\alpha}^2/2 M_{\alpha} -
{k_{\alpha}^{0}}^2/2 \mu_{\alpha}]}
 \int \frac{d {\rm{\bf{q}}}_{\alpha}^{0} }{(2\pi)^{3}}
e^{i {\rm{\bf{q}}}_{\alpha}^{0} \cdot
({\mbox {\boldmath $\rho$}}_{\alpha}'-
{\mbox {\boldmath $\rho$}}_{\alpha})} \nonumber \\
&=& \lim_{\delta \to +0}
\int \frac{d {\rm{\bf{k}}}_{\alpha}^{0}}{(2\pi)^{3}}
\int_{{\rho}_{\alpha} \geq A}
d {\mbox {\boldmath $\rho$}}_{\alpha}
\frac{e^{-i {\rm{\bf{\Delta}}}_{\alpha}\cdot
{\mbox {\boldmath $\rho$}}_{\alpha}}}{{\rho}_{\alpha}^{4}}
e^{-2\delta {\rho}_{\alpha}} \nonumber \\
&&\times \frac{D_{m {\rm{\bf{k}}}_{\alpha}^{0}
({\mbox {\boldmath $\rho$}}_{\alpha})}
(\hat {\mbox {\boldmath $\rho$}}_{\alpha})
D_{{\rm{\bf{k}}}_{\alpha}^{0}({\mbox{\boldmath$\rho$}}_{\alpha}) m}
(\hat {\mbox {\boldmath $\rho$}}_{\alpha})}
{[E + i0 - q_{\alpha}^2/2 M_{\alpha} -
{k_{\alpha}^{0}}^2/2 \mu_{\alpha}]}, \label{vcoo6a}
\end{eqnarray}
where $\bar {\rm{\bf{k}}}_{\nu}^0$ is given by the relation
(\ref{lc}), but with ${\rm{\bf{q}}}_{\alpha}$ instead of
${\rm{\bf{q}}}_{\alpha}^{0}$. Similarly, $\bar \eta_{\nu}^0$
is defined as in (\ref{cpar}), but with $\bar {\rm{\bf{k}}}_{\nu}^0$
instead of ${\rm{\bf{k}}}_{\nu}^0$.

We remark that the fact that the leading term of (\ref{vcoo6})
for ${\Delta}_{\alpha} \rightarrow 0$ is defined by that part
of the integration region where ${\mbox {\boldmath $\rho$}}_{\alpha}'
={\mbox {\boldmath $\rho$}}_{\alpha}$, was derived for an
analogous two-body problem in \cite{am94}. As is apparent,
the additional integration over ${\rm{\bf{k}}}_{\alpha}^{0}$,
which is present in the three-body case under
consideration, does not invalidate that result.

Next we observe that, since the behavior of
${\cal V}_{\alpha m, \alpha m\,(c)}^{opt\,(as)}$ for
${\Delta}_{\alpha} \rightarrow 0$ is defined by the behavior
of the integrand in (\ref{vcoo6a}) for
$\rho_{\alpha} \rightarrow \infty$, we can approximate
$D_{{\rm{\bf{k}}}_{\alpha}^{0}({\mbox {\boldmath $\rho$}}_{\alpha}) m}
(\hat {\mbox {\boldmath $\rho$}}_{\alpha})$ by
$D_{{\rm{\bf{k}}}_{\alpha}^{0} m}(\hat {\mbox {\boldmath
$\rho$}}_{\alpha})$, i.e. by the
same expression (\ref{dkm}) but with the local momentum
${\rm{\bf{k}}}_{\alpha}^{0}({\mbox {\boldmath $\rho$}}_{\alpha})$
replaced by its asymptotic value ${\rm{\bf{k}}}_{\alpha}^{0}$
which is attained in this limit. In fact,
taking into account (\ref{dkm}), (\ref{psis}) and (\ref{kloc})
one sees that
\begin{equation}
D_{{\rm{\bf{k}}}_{\alpha}^{0}({\mbox{\boldmath $\rho$}}_{\alpha}) m}
(\hat {\mbox {\boldmath $\rho$}}_{\alpha}) =
D_{{\rm{\bf{k}}}_{\alpha}^{0} m}(\hat {\mbox {\boldmath
$\rho$}}_{\alpha}) + O\left(\frac{1}{\rho_{\alpha}}\right). \label{das}
\end{equation}
An analogous expansion applies to
$D_{m {\rm{\bf{k}}}_{\alpha}^{0}({\mbox {\boldmath
$\rho$}}_{\alpha})}(\hat {\mbox {\boldmath $\rho$}}_{\alpha})$.
It is important to realize that (\ref{das}) is
valid for any $k_{\alpha}^0 > 0$: for arbitrary
positive-definite $k_{\alpha}^0$ always such a large $A$ can
be found that for all $\rho_{\alpha} \geq A$ the second term
in (\ref{das}) is infinitely small compared to the first one.
This obviously does no longer hold true for ${k}_{\alpha}^{0} = 0$,
as follows from the definition (\ref{kloc})
of the local momentum. But the whole integral is not
influenced by the behavior of the integrand in an
infinitesimal vicinity of ${k}_{\alpha}^{0}=0$ if the latter
does not possess a nonintegrable singularity there.
This is verified in Appendix \ref{appa} for
$q_{\alpha} \neq \sqrt{2 M_{\alpha}E}$. Thus we
are justified in using the approximation (\ref{das}) in the
whole region of integration over ${\rm {\bf {k}}}_{\alpha}^{0}$
including the origin, thereby neglecting the terms of the
order $O(1/\rho_{\alpha})$.

Introducing (\ref{das}) in (\ref{vcoo6a}) yields in
the leading order
\begin{eqnarray}
{\cal V}_{\alpha m, \alpha m\,(c)}^{opt\,(as)}
({\rm{\bf{q}}}_{\alpha}', {\rm{\bf{q}}}_{\alpha}; E)
&\stackrel{{\Delta}_{\alpha} \rightarrow 0}{\approx}&
\int_{{\rho}_{\alpha} \geq A} d {\mbox {\boldmath $\rho$}}_{\alpha}
\frac{e^{-i {\rm{\bf{\Delta}}}_{\alpha} \cdot
{\mbox {\boldmath $\rho$}}_{\alpha}}}{{\rho}_{\alpha}^{4}}
\int \frac {d {\rm{\bf{k}}}_{\alpha}^{0}}{(2 \pi)^3}
\frac{|D_{{\rm{\bf{k}}}_{\alpha}^{0} m}
(\hat {\mbox {\boldmath $\rho$}}_{\alpha})|^2}
{[E + i0 -q_{\alpha}^2/2 M_{\alpha} -
{k_{\alpha}^{0}}^2/2 \mu_{\alpha}]}\label{vcoo7}
\end{eqnarray}
(again, because of the convergence factor $\rho_{\alpha}^{-4}$
the limit $\delta \to +0$ could be performed before integrating
over ${\mbox {\boldmath $\rho$}}_{\alpha}$).

Let us write, in analogy to (\ref{dnm1}),
\begin{eqnarray}
D_{{\rm{\bf{k}}}_{\alpha}^{0} m}(\hat {\mbox {\boldmath $\rho$}}_
{\alpha}) &=& \frac {i}{\rho_{\alpha}} \lim_{{\rm{\bf{p}}} \to 0}
({\rm{\bf{D}}}_{{\rm{\bf{k}}}_{\alpha}^{0} m}
\cdot {\mbox {\boldmath $\nabla$}}_{\rm{\bf{p}}})
e^{-i {\rm{\bf{p}}} \cdot {\mbox {\boldmath $\rho$}}_{\alpha}},
\label{dkm1}
\end{eqnarray}
with
\begin{eqnarray}
{\rm{\bf{D}}}_{{\rm{\bf{k}}}_{\alpha}^{0} m}
= \epsilon_{\alpha \beta} e_{\alpha}
(\lambda_{\beta}e_{\gamma} - \lambda_{\gamma}e_{\beta})
\int d {\rm{\bf{r}}}_{\alpha} \psi_{{\rm{\bf{k}}}_{\alpha}^{0}}^{(+)*}
({\rm{\bf{r}}}_{\alpha}) {\rm{\bf{r}}}_{\alpha}
\psi_{\alpha m}({\rm{\bf{r}}}_{\alpha}),
\label{dkm2}
\end{eqnarray}
and similarly for $D_{m {\rm{\bf{k}}}_{\alpha}^{0}}
(\hat {\mbox {\boldmath $\rho$}}_{\alpha})$ and
${\rm{\bf{D}}}_{m {\rm{\bf{k}}}_{\alpha}^{0}}$
(cf. (\ref{sym})). Then we obtain from (\ref{vcoo7})
\begin{eqnarray}
{\cal V}_{\alpha m, \alpha m\,(c)}^{opt\,(as)}
({\rm{\bf{q}}}_{\alpha}', {\rm{\bf{q}}}_{\alpha}; E)
\stackrel{{\Delta}_{\alpha} \rightarrow 0}{\approx}
\lim_{{\rm{\bf{p}}},{\rm{\bf{p}}}' \to 0}
\int \frac{d {\rm{\bf{k}}}_{\alpha}^{0}}{(2\pi)^{3}}
\frac{({\rm{\bf{D}}}_{m {\rm{\bf{k}}}_{\alpha}^{0}}
\cdot {\mbox {\boldmath $\nabla$}}_{\rm{\bf{p}}'})
({\rm{\bf{D}}}_{{\rm{\bf{k}}}_{\alpha}^{0} m} \cdot
{\mbox {\boldmath $\nabla$}}_{\rm{\bf{p}}}) J(\rm{\bf{u}})}
{[E + i0 -q_{\alpha}^2/2 M_{\alpha} -
{k_{\alpha}^{0}}^2/2 \mu_{\alpha}]}.
\label{vcoo8}
\end{eqnarray}
The integral $J(\rm{\bf{u}})$ is given by (\ref{ju}).
Making use of its asymptotic form (\ref{b3}) for $u \to 0$ we find
\begin{eqnarray}
{\cal V}_{\alpha m, \alpha m\,(c)}^{opt\,(as)}
({\rm{\bf{q}}}_{\alpha}', {\rm{\bf{q}}}_{\alpha}; E)
\stackrel{{\Delta}_{\alpha} \rightarrow 0}{=}
C_{m}^{(c)} {\Delta}_{\alpha} + o\,(\Delta_{\alpha}),
\label{vas4}
\end{eqnarray}
with
\begin{eqnarray}
C_{m}^{(c)} = -\frac{\pi^2}{4}\int
\frac{d {\rm{\bf{k}}}_{\alpha}^{0}}{(2\pi)^{3}}
\frac{\left[|{\rm{\bf{D}}}_{{\rm{\bf{k}}}_{\alpha}^{0} m}|^2 +
|\hat {\rm {\bf {\Delta}}}_{\alpha} \cdot
{\rm{\bf{D}}}_{{\rm{\bf{k}}}_{\alpha}^{0} m}|^2 \right]}
{[E + i0 -q_{\alpha}^2/2 M_{\alpha} -
{k_{\alpha}^{0}}^2/2 \mu_{\alpha}]}. \label{ck}
\end{eqnarray}
As in (\ref{vas3}), we omitted also in (\ref{vas4}) a
${\Delta}_{\alpha}$-independent term. We point out that from
the discussion in the Appendix \ref{appa} follows that the
result (\ref{vas4}) with (\ref{ck}) is valid also for
$q_{\alpha}=\sqrt{2 M_{\alpha}E}$ where, in contrast to
appearance, $C_{m}^{(c)}$ in (\ref{ck}) is not singular.

For on-shell scattering ($q_{\alpha}'=q_{\alpha}=
\bar q_{\alpha}$), taking into account (\ref{dk1})
the expression (\ref{ck}) simplifies to
\begin{eqnarray}
C_{m}^{(c)} = \frac{\pi^2}{4} \int
\frac{d {\rm{\bf{k}}}_{\alpha}^{0}}{(2\pi)^{3}}
\frac{\left[|{\rm{\bf{D}}}_{{\rm{\bf{k}}}_{\alpha}^{0} m}|^2 +
|\hat {\rm {\bf {\Delta}}}_{\alpha} \cdot
{\rm{\bf{D}}}_{{\rm{\bf{k}}}_{\alpha}^{0} m}|^2 \right]}
{\left[|\hat E_{\alpha m}| + {k_{\alpha}^{0}}^2
/2 \mu_{\alpha}\right]}. \label{ck1}
\end{eqnarray}
That is, in the leading order ${\cal V}_{\alpha m,
\alpha m\,(c)}^{opt\,(as)}$ depends neither on the incoming
nor the outgoing momenta ${\rm{\bf{q}}}_{\alpha}$ and
${\rm{\bf{q}}}_{\alpha}'$ alone but only on the momentum
transfer ${\rm{\bf{\Delta}}}_{\alpha}$. It is also independent
of the energy. Hence we can write (\ref{vcoo7}) as
\begin{eqnarray}
{\cal V}_{\alpha m, \alpha m\,(c)}^{opt\,(as)}
({\rm{\bf{q}}}_{\alpha}', {\rm{\bf{q}}}_{\alpha}; E_{\alpha m}+i0)
&\equiv& {\cal V}_{\alpha m, \alpha m\,(c)}^{opt\,(as)}
({\rm {\bf {\Delta}}}_{\alpha}) \nonumber \\
&\stackrel{{\Delta}_{\alpha} \rightarrow 0}{\approx}&
\int_{{\rho}_{\alpha} \geq A}d {\mbox {\boldmath $\rho$}}_{\alpha}
e^{-i {\rm {\bf {\Delta}}}_{\alpha} \cdot
{\mbox {\boldmath $\rho$}}_{\alpha}}
{\cal V}_{\alpha m, \alpha m\,(c)}^{opt\,(as)}
({\mbox {\boldmath $\rho$}}_{\alpha}), \label{vdeltc}
\end{eqnarray}
where the local potential
\begin{eqnarray}
{\cal V}_{\alpha m, \alpha m\,(c)}^{opt\,(as)}
({\mbox {\boldmath $\rho$}}_{\alpha}) =
\left\{ \int\frac{d {\rm{\bf{k}}}_{\alpha}^{0}}{(2\pi)^{3}}
\frac{|D_{{\rm{\bf{k}}}_{\alpha}^{0} m}
(\hat {\mbox {\boldmath $\rho$}}_{\alpha})|^2 }
{[|\hat E_{\alpha m}| + {k_{\alpha}^{0}}^2/2 \mu_{\alpha}]}
\right\} \frac{1}{\rho_{\alpha}^4} \label{vascooc}
\end{eqnarray}
describes the coordinate-space asymptotics of
${\cal V}_{\alpha m, \alpha m\,(c)}^{opt\,(as)}$. It is
the contribution to the polarisation potential
from the intermediate three-body continuum states.

Summing up the contribution (\ref{vas3}) with (\ref{cn})
from the intermediate two-fragment states and the
contribution (\ref{vas4}) with (\ref{ck}) from the
intermediate three-particle continuum states, we thus have
off the energy shell the final result
\begin{eqnarray}
{\cal V}_{\alpha m, \alpha m}^{opt\,(as)}
({\rm{\bf{q}}}_{\alpha}', {\rm{\bf{q}}}_{\alpha}; E)
\stackrel{{\Delta}_{\alpha} \rightarrow 0}{=}
C_{1} \Delta_{\alpha} + o\,(\Delta_{\alpha}), \label{vas5}
\end{eqnarray}
with
\begin{eqnarray}
C_{1}&=& C_{m}^{(b)} + C_{m}^{(c)} \nonumber \\
&=& -\frac{\pi^2}{4}\left\{\sum_{n \neq m}
\frac{\left[|{\rm{\bf{D}}}_{nm}|^2 +
|\hat {\rm {\bf {\Delta}}}_{\alpha}
\cdot {\rm{\bf{D}}}_{nm}|^2 \right]}
{[E +i0 - q_{\alpha}^2/2 M_{\alpha} - \hat E_{\alpha n}]}
+ \int \frac{d {\rm{\bf{k}}}_{\alpha}^{0}}{(2\pi)^{3}}
\frac{\left[|{\rm{\bf{D}}}_{{\rm{\bf{k}}}_{\alpha}^{0} m}|^2 +
|\hat {\rm {\bf {\Delta}}}_{\alpha} \cdot
{\rm{\bf{D}}}_{{\rm{\bf{k}}}_{\alpha}^{0} m}|^2 \right]}
{[E + i0 -q_{\alpha}^2/2 M_{\alpha} -
{k_{\alpha}^{0}}^2/2 \mu_{\alpha}]}\right\}. \label{cmm}
\end{eqnarray}

Consequently, for off-shell scattering we have found
for the asymptotic
behavior of the total optical potential (\ref{mvopt}), i.e.
including the contribution from the static potential (for
a spherically symmetric target state $m$), in the
limit ${\Delta}_{\alpha} \rightarrow 0$ the result
\begin{equation}
{\cal V}_{\alpha m, \alpha m}^{opt}
({\rm{\bf{q}}}_{\alpha}', {\rm{\bf{q}}}_{\alpha}; E)
\stackrel{{\Delta}_{\alpha} \rightarrow 0}{=}
\frac{4\pi e_{\alpha}(e_{\beta}+e_{\gamma})}
{{\Delta}_{\alpha}^2} + C_{1} \Delta_{\alpha} +
o\,(\Delta_{\alpha}), \label{vas5a}
\end{equation}
with the energy- and momentum-dependent factor $C_{1} =
C_{1}(q_{\alpha},E)$ given by (\ref{cmm}). It should be
noted that the strength factor $C_{1}$ of the leading term
decreases with increasing energy.

Equation (\ref{vas5a}) with (\ref{cmm}) constitutes our first
main result. It implies the exact compensation in the
(non-static part of the) optical potential of the singular
terms proportional to ${\Delta}_{\alpha}^{-1}$ and
$\ln{{\Delta}_{\alpha}}$, in contrast to their appearence
(as discussed in \cite{ks87} for negative energies) in the
effective potentials which occur in the
effective-two-body formulation of the three-charged particle
theory. We emphasize that this compensation
was proved here both on and off the energy shell, and for
arbitrary total three-body energy $E<0$ and $E>0$. It
generalizes the cancellation proofs,
given in \cite{km88,ksz85,ks87,ksp87,ks89,zep90} for
energies below the dissociation threshold, to positive
energies. In addition, at negative three-particle energy our
derivation avoids the various inconsistencies existing there.

If we consider on-shell scattering then, as follows from
(\ref{vas3}), (\ref{cn1}), (\ref{vas4}) and (\ref{ck1}),
$ {\cal V}_{\alpha m, \alpha m}^{opt\,(as)}$ depends on the
momenta only in the form of the momentum transfer
${\rm {\bf {\Delta}}}_{\alpha}$, and
no longer on the energy, i.e.,
\begin{equation}
 {\cal V}_{\alpha m, \alpha m}^{opt\,(as)}
({\rm{\bf{q}}}_{\alpha}', {\rm{\bf{q}}}_{\alpha};
E_{\alpha m}) \equiv
 {\cal V}_{\alpha m, \alpha m}^{opt\,(as)}
({\rm {\bf {\Delta}}}_{\alpha}). \label{voptoes}
\end{equation}
Consequently, the inverse Fourier transform of
$ {\cal V}_{\alpha m, \alpha m}^{opt\,(as)}
({\rm {\bf {\Delta}}}_{\alpha})$ is a local,
energy-independent potential describing the asymptotic
behavior of the (non-static part of the) optical
potential in coordinate space. From (\ref{vascoo}) and
(\ref{vascooc}) we read off
\begin{eqnarray}
 {\cal V}_{\alpha m, \alpha m}^{opt\,(as)}
({\mbox {\boldmath $\rho$}}_{\alpha})
\stackrel{\rho_{\alpha} \to \infty}{=}
- \frac{a}{2 \rho_{\alpha}^4}
+ o\left( \frac{1}{\rho_{\alpha}^4}\right), \label{coo}
\end{eqnarray}
with
\begin{eqnarray}
a=2 \sum_{n \neq m}\frac{|D_{nm}(\hat{\mbox{\boldmath $\rho$}}
_{\alpha})|^2}{|\hat E_{\alpha m}| - |\hat E_{\alpha n}|}
+2 \int \frac{d {\rm{\bf{k}}}_{\alpha}^{0}}
{(2\pi)^{3}}\frac{|D_{{\rm{\bf{k}}}_{\alpha}^{0} m}
(\hat {\mbox {\boldmath $\rho$}}_{\alpha})|^2 }
{|\hat E_{\alpha m}| + {k_{\alpha}^{0}}^2/2 \mu_{\alpha}}.
\label{adip}
\end{eqnarray}
As can be seen, the strength factor governing the
coordinate-space behavior is just the static dipole
polarisability as known from the perturbative approaches.
That is, no remormalization of $a$ arises from summing up
all the higher order terms in the perturbation expansion
of the full three-particle Green function $G(z)$. Furthermore,
all dependence on the incoming energy has disappeared
(in contrast to the approximate result of ref. \cite{mcc82},
but in agreement with the result obtained in \cite{uc89}
within the adiabatic approach).

For the coordinate-space behavior of the total optical
potential, i.e. the inverse Fourier transform of
(\ref{mvopt}), we
derive on the energy shell for spherically symmetric targets
\begin{eqnarray}
{\cal V}_{\alpha m, \alpha m}^{opt}
({\mbox{\boldmath $\rho$}}_{\alpha})
\stackrel{\rho_{\alpha} \longrightarrow \infty}{=}
\frac{e_{\alpha}(e_{\beta}+e_{\gamma})}{\rho_{\alpha}}-
\frac{a}{2 \rho_{\alpha}^4} + o\,\left( \frac{1}
{\rho_{\alpha}^4}\right). \label{mvopt1}
\end{eqnarray}
Equation (\ref{mvopt1}) with (\ref{adip}) is our second main
result. It states that in the asymptotic expansion of the
energy-shell restriction of the optical potential for large
intercluster separation there occurs as the first nonvanishing
term after the center-of-mass Coulomb interaction, which arises
from the multipole expansion of the folded Coulomb channel
interaction, the local (induced) polarisation potential
$-a/2\rho_{\alpha}^4$. This result extends the one derived
in \cite{km88,ksz85,ks87,ksp87,ks89,zep90} for $E<0$,
to arbitrary energies.

Simultaneously we derived the exact expression for the static
dipole polarisability $a$. To our best knowledge this is the
first derivation of $a$ within the framework of a genuinly
three-body, non-perturbative, non-adiabatic approach. In this
context we point out that the same expression
for $a$ was derived in
\cite{km88} for $E<0$. However, in that derivation a spectral
decomposition of the three-body Green function in
$\Omega_{\alpha} \cap \Omega_{\alpha}'$ was used,
which is inadequate since it does not take into account
the long-ranged three-body correlations described in
\cite{am92}. The latter necessarily lead to the appearence
of the wave functions $\psi^{(+)}_{{\rm{\bf{k}}}_{\alpha}
({\mbox {\boldmath $\rho$}}_{\alpha})}({\rm {\bf{r}}}_{\alpha})$
instead of $\psi^{(+)}_{{\rm{\bf{k}}}_{\alpha}}
({\rm {\bf{r}}}_{\alpha})$. Though, this inconsistency in
\cite{km88} did not influence the final result. Nevertheless,
we feel that it is important to present a consistent
three-body derivation of the dipole polarisability.

\newpage
\section{Summary} \label{summ}

We have investigated the singularity structure in momentum
space of the optical potential responsible for the elastic
scattering of a charged particle $\alpha$ off a
bound state of two charged (or one charged and one neutral)
particles, on the basis of the rigourous three-body
scattering theory. Without having recourse to perturbative
or other approximate methods we have shown that (for
spherically symmetric bound states) the
optical potential behaves, in the limit that the momentum
transfer $\Delta_{\alpha}$ goes to zero, as $C_0
\Delta_{\alpha}^{-2} + C_1 \Delta_{\alpha} +
o\,(\Delta_{\alpha})$, for both negative {\em and}
positive energies.
That is, terms $\sim \Delta_{\alpha}^{-1}$ or
$\sim \ln \Delta_{\alpha}$ cancel exactly. The factors
appearing in this expansion have the familiar physical
interpretation: $C_0$ is the product of the total
charges of the colliding fragments, and $C_1$ is the
general, energy- and momentum-dependent expression for
the polarisability of the composite particle.

For on-shell scattering we find that this momentum-transfer
behavior entails in coordinate space a local tail of the optical
potential of the form $C_0/\rho_{\alpha} - a/2\rho_{\alpha}^{4} +
o\,(\rho_{\alpha}^{-4})$, where $\rho_{\alpha}$ denotes
the distance of the elementary from the center of
mass of the composite particle. The polarisability
$a$ derived here coincides with the expression as
extracted in perturbative or similar approximate
approaches. These results imply that \\
(i) the exact cancellation of terms $\sim \rho_{\alpha}^{-2}$
and $\sim \rho_{\alpha}^{-3}$ takes place for energies
below and above the dissociation threshold, \\
(ii) no renormalization of the polarisability
$a$ due to summing up the infinite
perturbation series occurs, and \\
(iii) the strength factor $a$ which governs the long-distance
behavior in coordinate space is independent of the incident energy
(in contrast to its off-shell analog $C_1$ which determines the
strength of the corresponding singularity in momentum space).

\vspace{1,7cm}

\noindent
$^{\dag}$ Permanent address: Institute for Nuclear Physics,
Tashkent, Usbekistan \\
$^{*}$ Supported by the Deutsche Forschungsgemeinschaft,
Project no. 436 USB-113-1-0

\newpage

\setcounter{section}{1}
\setcounter{equation}{0}
\def\theequation{\Alph{section}.\arabic{equation}}
\begin{appendix}
\section{} \label{appa}

In this Appendix we first investigate the behavior in
the limit $\Delta_{\alpha} \equiv |{\rm{\bf{q}}}_{\alpha}'-
{\rm{\bf{q}}}_{\alpha}| \rightarrow 0$ of the
contribution of the $n$-th intermediate bound state
in expression (\ref{vcoo4}), to be denoted by
${\cal V}_{\alpha m, \alpha m\,(n)}^{opt\,(as)}$,
at the discrete point
$q_{\alpha} = q_{\alpha n}$. To simplify
the considerations we replace the Coulomb scattering wave
functions $\psi^{(+)*}_{{\rm{\bf{q}}}_{\alpha}^{0}}
({\mbox {\boldmath $\rho$}}_{\alpha})$ and
$\psi^{(+)}_{{\rm{\bf{q}}}_{\alpha}^{0}}
({\mbox {\boldmath $\rho$}}_{\alpha}')$ by their leading
terms for large distances, namely by the planes waves
$e^{-i {\rm{\bf{q}}}_{\alpha}^{0} \cdot
{\mbox {\boldmath $\rho$}}_{\alpha}}$ and
$e^{i {\rm{\bf{q}}}_{\alpha}^{0} \cdot
{\mbox {\boldmath $\rho$}}_{\alpha}'}$, respectively.
We also omit the functions
$D_{nm}(\hat {\mbox {\boldmath $\rho$}}_{\alpha})$
and $D_{mn}(\hat {\mbox {\boldmath $\rho$}}_{\alpha}')$. By
doing so the question of the integrability or
non-integrability of the singularities under considerations
is not influenced. We therefore consider
\begin{eqnarray}
{\cal V}_{\alpha m, \alpha m\,(n)}^{opt\,(as)}
({\rm{\bf{q}}}_{\alpha}', {\rm{\bf{q}}}_{\alpha}; E)
&\stackrel{{\Delta}_{\alpha} \rightarrow 0}{\sim}&
2 M_{\alpha}\lim_{\delta \to +0}
\int \frac{d {\rm{\bf{q}}}_{\alpha}^{0}}{(2\pi)^{3}}
\int_{{\rho}_{\alpha}'
\geq A}d {\mbox {\boldmath $\rho$}}_{\alpha}'
\int_{{\rho}_{\alpha} \geq A}
d {\mbox {\boldmath $\rho$}}_{\alpha}
\frac{e^{i ({\rm{\bf{q}}}_{\alpha}^{0} -{\rm{\bf{q}}}_{\alpha}') \cdot
{\mbox {\boldmath $\rho$}}_{\alpha}'}}{{\rho}_{\alpha}^{'2}}
\nonumber \\ && \times
e^{-\delta {\rho}_{\alpha}'}\frac{1}{[q_{\alpha n}^{2} + i0 -
{q_{\alpha}^{0}}^{2}]} e^{-\delta \rho_{\alpha}}
\frac{-e^{i ({\rm{\bf{q}}}_{\alpha}^{0}-{\rm{\bf{q}}}_{\alpha})
\cdot{\mbox {\boldmath $\rho$}}_{\alpha}}}{\rho_{\alpha}^2}
\nonumber \\
&\stackrel{{\Delta}_{\alpha} \rightarrow 0}{\sim}&
\lim_{\delta \to +0}
\int \frac{d {\rm{\bf{q}}}_{\alpha}^{0}}{(2\pi)^{3}}
\frac{1}{\sqrt{({\rm{\bf{q}}}_{\alpha}^{0}-
{\rm{\bf{q}}}_{\alpha}')^2 + \delta^2}}
\frac{1}{[q_{\alpha n}^{2} + i0 - {q_{\alpha}^{0}}^{2}]}
\nonumber \\
&&\times \frac{1}{\sqrt{({\rm{\bf{q}}}_{\alpha}^{0}-
{\rm{\bf{q}}}_{\alpha})^2 + \delta^2}}.\label{a1}
\end{eqnarray}
The integrals over $\rho_{\alpha}$ and $\rho_{\alpha}'$
from $0$ to $A$ added in the second line are certainly less
singular than the original expression; hence they do not
alter our conclusion. We use the integral representation
\begin{equation}
\frac{1}{\sqrt{\zeta}} = \frac{1}{2 \pi} \oint d\beta
\frac{1}{\sqrt{\beta}} \frac{1}{\beta + \zeta} =
\frac{1}{\pi} \int_0^{\infty} d\beta
\frac{1}{\sqrt{\beta}} \frac{1}{\beta + \zeta}, \label{a2}
\end{equation}
where $\oint$ means the integral along a closed contour around the
point $\beta = - \zeta$; in the integral from $0$ to
$\infty$ of the second representation we must have
$\arg{\beta} = 0$. This leads to
\begin{eqnarray}
{\cal V}_{\alpha m, \alpha m\,(n)}^{opt\,(as)}
({\rm{\bf{q}}}_{\alpha}', {\rm{\bf{q}}}_{\alpha}; E)
&\stackrel{{\Delta}_{\alpha} \rightarrow 0}{\sim}&
\lim_{\delta \to +0} \int_0^{\infty} d\beta
\frac{1}{\sqrt{\beta}}\int_0^{\infty} d\alpha
\frac{1}{\sqrt{\alpha}} J(\beta,\alpha;{\rm {\bf {q}}}_{\alpha}',
{\rm {\bf {q}}}_{\alpha},q_{\alpha n}), \label{a3}
\end{eqnarray}
with
\begin{eqnarray}
&&J(\beta,\alpha;{\rm {\bf {q}}}_{\alpha}',
{\rm {\bf {q}}}_{\alpha},q_{\alpha n}) = \nonumber \\
&&\int \frac{d {\rm{\bf{q}}}_{\alpha}^{0}}
{(2\pi)^{3}} \frac{1}{[\beta + ({\rm{\bf{q}}}_{\alpha}^{0}-
{\rm{\bf{q}}}_{\alpha}')^2 + \delta^2]}
\frac{1}{[q_{\alpha n}^{2} + i0 - {q_{\alpha}^{0}}^{2}]}
\frac{1}{[\alpha + ({\rm{\bf{q}}}_{\alpha}^{0}-
{\rm{\bf{q}}}_{\alpha})^2 + \delta^2]}.\label{a4}
\end{eqnarray}
This integral is explicitly known \cite{lew56},
\begin{eqnarray}
J(\beta,\alpha;{\rm {\bf {q}}}_{\alpha}',{\rm {\bf {q}}}_{\alpha},
q_{\alpha n}) = - \frac{1}{8 \pi} \frac{1}{\sqrt{c^2 - ab}}
\ln{\left[\frac{c + \sqrt{c^2 - ab}}
{c - \sqrt{c^2 - ab}}\right]},\label{a5}
\end{eqnarray}
with
\begin{eqnarray}
ab &=& [\Delta_{\alpha}^2 +(\sqrt{\alpha}+\sqrt{\beta})^2]
[q_{\alpha}^2 + (\sqrt{\alpha}-iq_{\alpha n})^2]
[q_{\alpha}^{'2} + (\sqrt{\beta}-iq_{\alpha n})^2],
\label{a6}\\
c &=& -i q_{\alpha n}[\Delta_{\alpha}^2 +
(\sqrt{\alpha}+\sqrt{\beta})^2] +
\sqrt{\beta}[q_{\alpha}^2 + \alpha - q_{\alpha n}^2] +
\sqrt{\alpha}[q_{\alpha}^{'2} + \beta-q_{\alpha n}^2]. \label{a7}
\end{eqnarray}
Because of the presence of the parameters $\alpha$
and $\beta$, the regularisation parameter $\delta$ is no
longer needed. We therefore have already put
${\delta}$ equal to zero in (\ref{a5}) - (\ref{a7}). For
$q_{\alpha} =q_{\alpha}'= q_{\alpha n}$, (\ref{a6}) and
(\ref{a7}) simplify to
\begin{eqnarray}
ab &=& \sqrt{\alpha}\sqrt{\beta}[\Delta_{\alpha}^2 +(\sqrt{\alpha}
+\sqrt{\beta})^2][\sqrt{\alpha}-2i q_{\alpha n}]
[\sqrt{\beta}-2i q_{\alpha n}], \label{a8} \\
c&=& -i q_{\alpha n}[\Delta_{\alpha}^2 +(\sqrt{\alpha}+\sqrt{\beta})^2] +
\sqrt{\alpha \beta}[\sqrt{\alpha}+\sqrt{\beta}]. \label{a9}
\end{eqnarray}

Substitution of $\alpha= \Delta_{\alpha}^2 v$ and
$\beta=\Delta_{\alpha}^2 t$ in (\ref{a8}) and (\ref{a9})
shows that in the limit $\Delta_{\alpha} \rightarrow 0$
\begin{equation}
ab \sim \Delta_{\alpha}^4, \quad c \sim \Delta_{\alpha}^2,
\quad \mbox{i.e., } \sqrt{c^2-ab} \sim \Delta_{\alpha}^2. \label{a10}
\end{equation}
Introducing (\ref{a5}) into (\ref{a3}), changing to the new
variables $v$ and $t$, and taking into account (\ref{a10}),
it is immediately seen that the main term of
${\cal V}_{\alpha m, \alpha m\,(n)}^{opt\,(as)}$ is
independent of $\Delta_{\alpha}$ in the limit
$\Delta_{\alpha} \rightarrow 0$ at the discrete point
$q_{\alpha} = q_{\alpha n}$, i.e., it is regular there. Hence
the same holds true also for the sum over all intermediate
bound state contributions
${\cal V}_{\alpha m, \alpha m\,(b)}^{opt\,(as)}$.

The other problem concerns the behavior of the intermediate
three-body continuum part (\ref{vcoo6}) if $q_{\alpha } =
q_{\alpha 1}$, where $q_{\alpha 1}=\sqrt {2 M_{\alpha}E}$.
In that case we are not {\em a priori} allowed to take out
the propagator $d_{k_{\alpha}^{0}}(q_{\alpha}^{0})$ from
under the integral over ${\rm{\bf{q}}}_{\alpha}^{0}$ at the
point $q_{\alpha}^0 = q_{\alpha}$.
To investigate this case consider the definition (\ref{l}) of
$L({\rm{\bf{q}}}_{\alpha}^{0};{\mbox {\boldmath $\rho$}}_{\alpha},
{\mbox {\boldmath $\rho$}}_{\alpha}')$. Since in the
limit $\Delta_{\alpha} \rightarrow 0$ the leading,
$\Delta_{\alpha}$-dependent term is defined by
${\mbox {\boldmath $\rho$}}_{\alpha} \approx
{\mbox {\boldmath $\rho$}}_{\alpha}'$, the exponential Coulomb
distorsion factors cancel. That is, it suffices to consider
\begin{eqnarray}
L({\rm{\bf{q}}}_{\alpha}^{0};{\mbox {\boldmath $\rho$}}_{\alpha},
{\mbox {\boldmath $\rho$}}_{\alpha}') \approx
\int \frac{d {\rm{\bf{k}}}_{\alpha}^{0} }{(2\pi)^{3}}
\frac{D_{m {\rm{\bf{k}}}_{\alpha}^{0}
({\mbox {\boldmath $\rho$}}_{\alpha}')}
(\hat {\mbox {\boldmath $\rho$}}_{\alpha}')
D_{{\rm{\bf{k}}}_{\alpha}^{0}({\mbox {\boldmath $\rho$}}_{\alpha}) m}
(\hat {\mbox {\boldmath $\rho$}}_{\alpha})}
{[E + i0 -{q_{\alpha}^{0}}^2/2 M_{\alpha} -
{k_{\alpha}^{0}}^2/2 \mu_{\alpha}]}. \label{a11}
\end{eqnarray}

If we perform the ${\rm{\bf{k}}}_{\alpha}^{0}$-integration,
the pole of $d_{k_{\alpha}^{0}}(q_{\alpha}^{0})$ can
give rise to a singularity of $L({\rm{\bf{q}}}_{\alpha}^{0};
{\mbox {\boldmath $\rho$}}_{\alpha},
{\mbox {\boldmath $\rho$}}_{\alpha}')$ in the
$q_{\alpha}^{0}$-plane at $q_{\alpha}^{0}= q_{\alpha 1}$.
It is an end point singularity which arises from the
coincidence of the propagator pole at $k_{\alpha}^{0}=
\sqrt{\mu_{\alpha}(q_{\alpha 1}^{2} - {q_{\alpha}^{0}}^{2})
/M_{\alpha}}$ with the lower limit
$ k_{\alpha}^{0}=0$ of integration over $k_{\alpha}^{0}$. Since
the leading, $\Delta_{\alpha}$-dependent term of
${\cal V}_{\alpha m, \alpha m\,(c)}^{opt \,(as)}$ in the limit
$\Delta_{\alpha} \to 0$ is generated by the coincidence of
the singularities of the integrand in the integral (\ref{vcoo6})
over ${\rm {\bf {q}}}_{\alpha}^{0}$ at the point
${\rm {\bf {q}}}_{\alpha}^{0}= {\rm {\bf {q}}}_{\alpha}=
{\rm {\bf {q}}}_{\alpha}^{'}$, the
appearence of an additional singularity, namely in the
the function $L$, will influence the leading term of
${\cal V}_{\alpha m, \alpha m\,(c)}^{opt\,(as)}$.

To investigate the analytic behavior of $L({\rm {\bf {q}}}_
{\alpha}^{0};{\mbox {\boldmath $\rho$}}_{\alpha},
{\mbox {\boldmath $\rho$}}_{\alpha}')$ for $q_{\alpha}^{0}
- q_{\alpha 1} \to 0$, consider the integrand in
(\ref{a11}). The denominator has a zero that can cause
the singularity under consideration. The behavior
for $k_{\alpha}^{0} \to 0$ of the numerator (cf. (\ref{dkm}))
is defined by the wave function $\psi^{(+)}_{{\rm{\bf{k}}}
_{\alpha}^{0}({\mbox{\boldmath $\rho$}}_{\alpha})}
({\rm {\bf{r}}}_{\alpha})$ which we write as
\begin{eqnarray}
\psi^{(+)}_{{\rm{\bf{k}}}_{\alpha}^{0}
({\mbox{\boldmath $\rho$}}_{\alpha})}({\rm {\bf{r}}}_{\alpha}) &=&
\text{exp}\left[\frac{{\rm{\bf{a}}}_{\alpha}
({\hat{\mbox{\boldmath $\rho$}}}_{\alpha})}{\rho_{\alpha}}
\cdot {\mbox {\boldmath $\nabla$}}_{{\rm{\bf{k}}}_{\alpha}^{0}}
\right]\psi^{(+)}_{{\rm {\bf{k}}}_{\alpha}^{0}}({\rm {\bf{r}}}_
{\alpha}) \nonumber\\
&{\approx}& \psi^{(+)}_{{\rm {\bf{k}}}_{\alpha}^{0}}
({\rm {\bf {r}}}_{\alpha}) \,+ \, \frac{{\rm{\bf{a}}}_{\alpha}
({\hat {\mbox {\boldmath $\rho$}}}_{\alpha})
\cdot {\mbox {\boldmath $\nabla$}}_{{\rm{\bf{k}}}_{\alpha}^{0}}}
{\rho_{\alpha}}
\psi^{(+)}_{{\rm {\bf{k}}}_{\alpha}^{0}}({\rm {\bf {r}}}_{\alpha}) +
O\left({\frac{1}{\rho_{\alpha}^2}}\right).\label{a12}
\end{eqnarray}
The threshold behavior for $k_{\alpha}^{0} \to 0$ of the
scattering wave function $\psi^{(+)}_{{\rm {\bf{k}}}_{\alpha}^{0}}
({\rm {\bf {r}}}_{\alpha})$ in the potential $V_{\alpha} =
V_{\alpha}^{S} + V_{\alpha}^{C}$ is entirely given by the
corresponding behavior of the Coulomb scattering wave function
$\psi^{(+)}_{C,{\rm {\bf{k}}}_{\alpha}^{0}}({\rm {\bf {r}}}_
{\alpha})$. This follows from the wellknown fact that
the threshold behavior of the radial wave
functions for two charged particles is the same for all
the partial waves and is defined by the factor \cite{baz}
\begin{eqnarray}
\mid N_{\alpha}\mid = \left[{\frac{2\pi \eta_{\alpha}^{0}}
{e^{2\pi \eta_{\alpha}^{0}} - 1}}\right]^{\frac{1}{2}}.
\label{a13a}
\end{eqnarray}
Therefore, the behavior of the full wave function
$\psi^{(+)}_{{\rm {\bf{k}}}_{\alpha}^{0}}
({\rm {\bf {r}}}_{\alpha})$ for $k_{\alpha}^{0} \to 0$
is governed by the same factor, and the one of the
numerator in (\ref{a11}) by ${\mid N_{\alpha} \mid}^{2}$.

Hence we have to distinguish two cases.

1. The Coulomb potential $V_{\alpha}^{C}$ is repulsive, i.e.,
$\eta_{\alpha}^{0} > 0$. Then for
$k_{\alpha}^{0} \to 0$ one has ${\mid N_{\alpha} \mid}^{2} \,
\approx \, 2\pi \eta_{\alpha}^{0} e^{-2\pi \eta_{\alpha}^{0}}
\to 0$. The exponential smallness of the
numerator for $k_{\alpha}^{0} \to 0$ compensates any
zero of the denominator at the point $k_{\alpha}^{0} = 0$.
In other words, the pole of the propagator in (\ref{a11})
at $k_{\alpha}^{0}= \sqrt{\mu_{\alpha}(q_{\alpha 1}^{2} -
{q_{\alpha}^{0}}^{2})/M_{\alpha}}$ does not generate a singularity of $L({\rm
{\bf {q}}}_{\alpha}^{0};
{\mbox {\boldmath $\rho$}}_{\alpha},
{\mbox {\boldmath $\rho$}}_{\alpha}')$ at the point
$q_{\alpha}^{0} = q_{\alpha 1}$. Thus,
when extracting the leading term of
${\cal V}_{\alpha m, \alpha m\,(c)}^{opt\,(as)}$ for
${\Delta}_{\alpha} \rightarrow 0$, the function
$L({\rm {\bf {q}}}_{\alpha}^{0};{\mbox {\boldmath $\rho$}}_{\alpha},
{\mbox {\boldmath $\rho$}}_{\alpha}')$ can be taken out
from under the integral over ${\rm{\bf{q}}}_{\alpha}^{0}$ at
the point ${\rm{\bf{q}}}_{\alpha}^{0}={\rm{\bf{q}}}_{\alpha}$,
as in the case $q_{\alpha} \neq q_{\alpha1}$.
The consequence is that the behavior of the optical potential
is the same for all $q_{\alpha}$ including $q_{\alpha} =
q_{\alpha1}$, namely $\sim O(\Delta_{\alpha})$.

2. The Coulomb potential $V_{\alpha}^{C}$ is attractive,
i.e., $\eta_{\alpha}^{0} < 0$.
In this case, ${\mid N_{\alpha} \mid}^{2} {\approx} 2\pi
\eta_{\alpha}^{0} \to \infty$ for $k_{\alpha}^{0} \to 0$,
and hence all terms in the expansion (\ref{a12})
become singular. This difficulty can
be overcome as follows. According \cite{am92} the wave function
$\Psi^{(+)as}_{{\rm{\bf{q}}}_{\alpha}^{0} {\rm{\bf{k}}}_{\alpha}^{0}}
({\mbox {\boldmath $\rho$}}_{\alpha},{\rm{\bf{r}}}_{\alpha})$, as
given by (\ref{psias}), is asymptotic solution of the three-particle
Schr\"odinger equation in the region $\Omega_{\alpha}$, i.e.,
it satisfies this equation up to terms $O(1/{\rho_{\alpha}^{2}})$
(cf. (\ref{seas})).
{}From the derivation given there it is, however, easily seen
that for $k_{\alpha}^{0}=0$, taking into account the local momentum
${\rm{\bf{k}}}_{\alpha}^{0}({\mbox {\boldmath $\rho$}}_{\alpha})$
in the asymptotic solution
$\Psi^{(+)as}_{{\rm{\bf{q}}}_{\alpha}^{0} {\rm{\bf{k}}}_{\alpha}^{0}}
({\mbox {\boldmath $\rho$}}_{\alpha},{\rm{\bf{r}}}_{\alpha})$
instead of its asymptotic value ${\rm {\bf {k}}}_{\alpha}^{0}=0$,
gives corrections
$\sim O(1/{\rho_{\alpha}^{2}})$ to the Schr\"odinger equation in
$\Omega_{\alpha}$. Therefore, in a small vicinity of
$k_{\alpha}^{0}=0$ the asymptotic solution
$\Psi^{(+)as}_{{\rm{\bf{q}}}_{\alpha}^{0} {\rm{\bf{k}}}_{\alpha}^{0}}
({\mbox {\boldmath $\rho$}}_{\alpha},{\rm{\bf{r}}}_{\alpha})$
may be replaced by the function
$\psi^{(+)}_{{\rm{\bf{k}}}_{\alpha}^{0}}
({\rm {\bf{r}}}_{\alpha})\,e^{ i {\rm{\bf{q}}}_{\alpha}^{0} \cdot
{\mbox {\boldmath $\rho$}}_{\alpha}}
e^{i \eta_{\beta}^{0} ln(k_{\beta}^{0} \rho_{\alpha}-
\epsilon_{\alpha \beta} {\rm{\bf{k}}}_{\beta}^{0} \cdot
{\mbox {\boldmath $\rho$}}_{\alpha})}
e^{i \eta_{\gamma}^{0} ln(k_{\gamma}^{0} \rho_{\alpha}-
\epsilon_{\alpha \gamma} {\rm{\bf{k}}}_{\gamma}^{0} \cdot
{\mbox {\boldmath $\rho$}}_{\alpha})}$, i.e., by
(\ref{psias}) but with ${\rm {\bf {k}}}_{\alpha}^{0}$
substituting ${\rm{\bf{k}}}_{\alpha}^{0}
({\mbox {\boldmath $\rho$}}_{\alpha})$ in
$\psi^{(+)}_{{\rm{\bf{k}}}_{\alpha}^{0}
({\mbox {\boldmath $\rho$}}_{\alpha})}({\rm {\bf{r}}}_{\alpha})$.
Hence for arbitrary $k_{\alpha}^{0}$ we may rewrite the asymptotic solution
$\Psi^{(+)as}_{{\rm{\bf{q}}}_{\alpha}^{0} {\rm{\bf{k}}}_{\alpha}^{0}}
({\mbox {\boldmath $\rho$}}_{\alpha},{\rm{\bf{r}}}_{\alpha})$
in the form (\ref{psias}) with
\begin{eqnarray}
{\rm{\bf{k}}}_{\alpha}^{0}({\mbox {\boldmath $\rho$}}_{\alpha})
&=& {\rm{\bf{k}}}_{\alpha}^0 + \chi(k_{\alpha}^{0})
\frac{{\rm{\bf{a}}}_{\alpha}({\hat {\mbox {\boldmath
$\rho$}}}_{\alpha})}{\rho_{\alpha}},\label{a14}
\end{eqnarray}
where $\chi(k_{\alpha}^{0})$ is a characteristic function,
which equals one everywhere except for a small neighbourhood
of $k_{\alpha}^{0}=0$, and goes smoothly to zero when
$k_{\alpha}^{0}$ approaches zero. Thus,
taking into account the factor ${k_{\alpha}^{0}}^{2}$
from the phase volume, the behavior of the numerator in
(\ref{a11}) in the limit $k_{\alpha}^{0} \to 0$ is governed by
${k_{\alpha}^{0}}^{2}{\mid N_{\alpha}\mid}^{2} \sim k_{\alpha}^{0}$.

Let us write $L({\rm {\bf {q}}}_{\alpha}^{0};{\mbox {\boldmath
$\rho$}}_{\alpha}, {\mbox {\boldmath $\rho$}}_{\alpha}')$ as the
sum $L = L_{\varepsilon} + \tilde{L_{\varepsilon}}$, where
$L_{\varepsilon}$ denotes the integral over
${\rm {\bf {k}}}_{\alpha}^{0}$ over the interior of a
small sphere with radius $\varepsilon$.
$\tilde{L_{\varepsilon}}$ contains the remaining
integral. Since the latter does not contain the origin
$k_{\alpha}^{0}=0$, $\tilde{L}_{\varepsilon}
({\rm {\bf {q}}}_{\alpha}^{0};{\mbox {\boldmath $\rho$}}_{\alpha},
{\mbox {\boldmath $\rho$}}_{\alpha}')$ is regular at
$q_{\alpha}^{0}=q_{\alpha1}=q_{\alpha}$ and
can therefore be taken out from under the integral over
${\rm{\bf{q}}}_{\alpha}^{0}$ in (\ref{vcoo6}) at the point
${\rm{\bf{q}}}_{\alpha}^{0}={\rm{\bf{q}}}_{\alpha}$. The
consequence is that at the momentum $q_{\alpha} = q_{\alpha1}$,
its contribution to the optical potential
$V_{\alpha}^{C}$ is $\sim O(\Delta_{\alpha})$,
i.e. of the same order as for $q_{\alpha} \neq q_{\alpha1}$.

However, $L_{\varepsilon}({\rm {\bf {q}}}_{\alpha}^{0};
{\mbox {\boldmath $\rho$}}_{\alpha},
{\mbox {\boldmath $\rho$}}_{\alpha}')$ is singular
at $q_{\alpha}^{0}=q_{\alpha1}$. When performing the
${\rm {\bf {k}}}_{\alpha}^{0}$-integration, the pole of
$d_{k_{\alpha}^{0}}(q_{\alpha}^{0})$ gives rise to a behavior
$L_{\varepsilon}({\rm {\bf {q}}}_{\alpha}^{0};
{\mbox {\boldmath $\rho$}}_{\alpha},
{\mbox {\boldmath $\rho$}}_{\alpha}')
\sim \ln({{q_{\alpha}^{0}}^{2} + i0 - q_{\alpha1}^{2}})$
in the $q_{\alpha}^{0}$-plane. In order to arrive at this
result we singled out from the product
$D_{m {\rm{\bf{k}}}_{\alpha}^{0}({\mbox {\boldmath $\rho$}}_{\alpha}')}
(\hat {\mbox {\boldmath $\rho$}}_{\alpha}')
D_{{\rm{\bf{k}}}_{\alpha}^{0}({\mbox {\boldmath $\rho$}}_{\alpha}) m}
(\hat {\mbox {\boldmath $\rho$}}_{\alpha})$ the factor
${\mid N_{\alpha} \mid}^{2}$ which is singular at
$k_{\alpha}^{0}=0$, and ignored the rest since it remains
finite. Denote by
${\cal V}_{\alpha m, \alpha m\,(\varepsilon)}^{opt\,(as)}
({\rm{\bf{q}}}_{\alpha}', {\rm{\bf{q}}}_{\alpha}; E)$
that part of ${\cal V}_{\alpha m, \alpha m \,(c)}^{opt\,(as)}
({\rm{\bf{q}}}_{\alpha}', {\rm{\bf{q}}}_{\alpha}; E)$ which
contains $L_{\varepsilon}({\rm {\bf {q}}}_{\alpha}^{0};
{\mbox {\boldmath $\rho$}}_{\alpha},
{\mbox {\boldmath $\rho$}}_{\alpha}')$
instead of $L({\rm {\bf {q}}}_{\alpha}^{0};
{\mbox {\boldmath $\rho$}}_{\alpha},
{\mbox {\boldmath $\rho$}}_{\alpha}')$ (cf. (\ref{vcoo6})).
Substituting $\ln({{q_{\alpha}^{0}}^{2} + i0 - q_{\alpha1}^{2}})$
for $L_{\varepsilon}({\rm {\bf {q}}}_{\alpha}^{0};
{\mbox {\boldmath $\rho$}}_{\alpha},
{\mbox {\boldmath $\rho$}}_{\alpha}')$, and
integrating over ${\mbox {\boldmath $\rho$}}_{\alpha}$
and ${\mbox {\boldmath $\rho$}}_{\alpha}'$, gives
\begin{eqnarray}
{\cal V}_{\alpha m, \alpha m\,(\varepsilon)}^{opt\,(as)}
({\rm{\bf{q}}}_{\alpha}', {\rm{\bf{q}}}_{\alpha}; E)
&\stackrel{{\Delta}_{\alpha} \rightarrow 0}{\sim}&
\lim_{\delta \to +0}
\int \frac{d {\rm{\bf{q}}}_{\alpha}^{0}}{(2\pi)^{3}}
\frac{\ln{(q_{\alpha 1}^{2}+i0-{q_{\alpha}^{0}}^{2})}}
{\sqrt{({\rm{\bf{q}}}_{\alpha}^{0}-
{\rm{\bf{q}}}_{\alpha}')^2 + \delta^2}
\sqrt{({\rm{\bf{q}}}_{\alpha}^{0}-
{\rm{\bf{q}}}_{\alpha})^2 + \delta^2}}.\label{a15}
\end{eqnarray}
When writing down this expression we already took into
account that the behavior of
${\cal V}_{\alpha m, \alpha m\,(\varepsilon)}^{opt\,(as)}
({\rm{\bf{q}}}_{\alpha}', {\rm{\bf{q}}}_{\alpha}; E)$ for
$\Delta_{\alpha} \to 0$ is defined by the region
${\mbox {\boldmath $\rho$}}_{\alpha}{\approx}
{\mbox {\boldmath $\rho$}}_{\alpha}'$ where the product
of the Coulomb distortion factors equals one. The integration
region over ${\rm {\bf {q}}}_{\alpha}^{0}$ in (\ref{a15})
supposedly consists only of the neighbourhood of
${\rm{\bf{q}}}_{\alpha}^{0} \approx {\rm {\bf {q}}}_{\alpha}
\approx {\rm {\bf {q}}}_{\alpha}'$. Comparing
with (\ref{a1}) (putting there $q_{\alpha n}$ equal to
$q_{\alpha 1}$) we conclude that expression (\ref{a15})
is less singular at
$\Delta_{\alpha} \to 0$ than (\ref{a1}). Hence,
in the case of an attractive Coulomb potential $V_{\alpha}^C$,
for $q_{\alpha}=q_{\alpha 1}$ the contribution
${\cal V}_{\alpha m, \alpha m\,(\varepsilon)}^{opt\,(as)}
({\rm{\bf{q}}}_{\alpha}', {\rm{\bf{q}}}_{\alpha}; E)$,
and thus also ${\cal V}_{\alpha m, \alpha m \,(c)}^{opt\,(as)}
({\rm{\bf{q}}}_{\alpha}', {\rm{\bf{q}}}_{\alpha}; E)$, is
finite in the limit $\Delta_{\alpha} \to 0$.

\newpage

\setcounter{equation}{0}
\section{} \label{appb}

In this Appendix we evaluate the integral (\ref{ju}) and extract
its asymptotic behavior for $u \to 0$. Substituting the new
variable $v = u \rho_{\alpha}$ and making use of eqs.
3.761(2) and 8.352(3) of \cite{gr65} we obtain
\begin{eqnarray}
J(u) &=& 2 \pi i u^3 [\Gamma(-4,iuA) - \Gamma(-4,-iuA)] = \nonumber \\
&=& \frac{i \pi u^3}{12} [\Gamma(0,iuA) - \Gamma(0,-iuA)] -
\sum_{m=0}^3 \frac{m!}{(i u A)^{m+1}}\left( e^{i u A} +
(-1)^m e^{-i u A} \right), \label{b1}
\end{eqnarray}
where $\Gamma(\mu,x)$ is the incomplete Gamma function \cite{gr65}.
With the help of eqs. 3.761(2), 8.230(1) and
8.232(1) of \cite{gr65} one finds
\begin{equation}
\Gamma(0,iuA) - \Gamma(0,-iuA)= 2 i \,\mbox{si}(uA) = -i\pi +
\sum_{m=0}^{\infty} \frac{(-1)^{m+1}(uA)^{2m-1}}{(2m-1)(2m-1)!},
\label{b2}
\end{equation}
where $\mbox{si}(x)$ is the sine-integral.
Substitution of (\ref{b2}) into (\ref{b1}) leads to the following
asymptotic behavior of $J(u)$ for $u \to 0$:
\begin{eqnarray}
J(u) \stackrel{u \to 0}{=} \frac{4\pi}{3A^3} - \frac{2\pi}{3A}u^2
+ \frac{\pi^2}{12} u^3 + O(u^4). \label{b3}
\end{eqnarray}

\end{appendix}

\end{document}